# A semi-analytic model for effects of fuzzy dark matter granule perturbations on orbital motion

Yu Zhao [1,2]★, Andrew Benson [2]★ and Xiaolong Du [3]★

[1] *Department of Physics and Astronomy, University of Southern California, Los Angeles, CA 90007, USA*
[2] *Carnegie Observatories, 813 Santa Barbara Street, Pasadena, CA 91101, USA*
[3] *Department of Physics and Astronomy, University of California, Los Angeles, CA 90095, USA*

## ABSTRACT

In fuzzy dark matter scenarios, the quantum wave nature of ultralight axion-like particles generates stochastic density fluctuations inside dark matter haloes. These fluctuations, known as granules, perturb the orbits of subhaloes and other orbiting bodies. While previous studies have simulated these effects using *N*-body techniques or modelled them statistically using diffusion approximations, we propose an alternative framework based on representing the perturbations as a Fourier series with random coefficients, which can be applied to individual orbits, not just populations. We extend the model to finite-size subhaloes, identifying a critical length scale below which subhaloes behave as point-mass particles. In contrast, larger subhaloes exhibit suppressed perturbations from granules due to their extended mass profiles. Using FDM-SIMULATOR, we validate our finite-size model by isolating granule accelerations and confirming their statistical effects on subhalo dynamics.

## 1 INTRODUCTION

The standard Λcold dark matter (ΛCDM) model has been remarkably successful in explaining the universe's large-scale structure, from galaxy clustering to cosmic microwave background anisotropies. However, it faces persistent challenges on galactic and subgalactic scales. These include the cusp-core problem, where ΛCDM simulations predict steep central density profiles in dark matter haloes, conflicting with observations of flat cores in dwarf galaxies (Flores & Primack 1994; Moore 1994; de Blok 2010); the missing satellites problem, which originally highlighted the discrepancy between the abundance of predicted subhaloes and observed dwarf galaxies (Klypin et al. 1999; Moore et al. 1999) but is now largely considered resolved, as recent studies suggest that ΛCDM may predict a deficit of satellites rather than an excess (Kim, Peter & Hargis 2018; Müller et al. 2024), and the too-big-to-fail problem, where simulated massive subhaloes are too dense to host observed bright satellites (Boylan-Kolchin, Bullock & Kaplinghat 2011; Tollerud, Boylan-Kolchin & Bullock 2014).

Since the success of CDM is exceptionally well established on linear scales, it follows that any alternative model of dark matter needs to match the behaviour of CDM on large scales in order to constitute a viable model of structure formation. One such alternative is fuzzy dark matter (FDM), which is composed of non-interacting ultralight bosons with masses $m_{\rm b} \sim 10^{-22}\,{\rm eV}$ (Hu, Barkana & Gruzinov 2000; Hui et al. 2017). These particles exhibit coherent dynamics and a wave-like behaviour on galactic scales due to their large de Broglie wavelength, which naturally suppresses small-scale structure formation. This suppression offers a potential solution to the small-scale problems and could explain the observed cored density profiles in dwarf galaxies (Schive, Chiueh & Broadhurst 2014a; Marsh & Pop 2015; Robles et al. 2015; Schive et al. 2016; May & Springel 2021). At larger scales, where the de Broglie wavelength is much smaller than the relevant structures, FDM retains the successes of CDM (Du, Behrens & Niemeyer 2017; Mocz et al. 2017; Zhang, Liu & Chu 2018). While firm limits have now been placed on models in which FDM makes up the entirety of the dark matter – from observations including the Ly α forest (Rogers & Peiris 2021), strong gravitational lensing (Laroche et al. 2022; Powell et al. 2023), and galaxy abundances (Kulkarni & Ostriker 2022) – models in which FDM makes up only a fraction of the total may remain viable. Exploring such scenarios will require fast and flexible approaches to modelling FDM physics.

The quantum nature of FDM gives rise to stable, self-gravitating solitonic cores at the centres of haloes (Schive et al. 2014a; Schwabe, Niemeyer & Engels 2016; Mocz et al. 2017; Zhang et al. 2018; Ferreira 2021). These solitons, which are non-linear ground-state solutions of the Schrödinger–Poisson (SP) equations with characteristic densities and radii set by $m_{\rm b}$, provide a natural explanation for the observed cores in dwarf galaxies, circumventing the cusp-core and too-big-to-fail problems of CDM (Robles, Bullock & Boylan-Kolchin 2019). FDM haloes form an extended envelope with a density profile similar to the Navarro–Frenk–White (NFW) profile seen in CDM haloes (Navarro, Frenk & White 1997) outside the soliton. Surrounding the solitonic core, the FDM halo is composed of excited quantum states whose interference generates persistent density fluctuations, or 'granules', on scales comparable

★ E-mail: yzhao112@usc.edu (YZ); abenson@carnegiescience.edu (AB); xdu@astro.ucla.edu (XD)

to the de Broglie wavelength (Schive et al. 2014a; Hui et al. 2017).

A typical FDM halo is characterized by order unity density fluctuations. These fluctuations arise from excited states, which produce quantum interference outside the solitonic core. Additionally, these excited modes affect the core itself (Li, Hui & Yavetz 2021), through both gravity and quantum pressure gradients, leading to oscillations and random motion that influence nearby structures. These fluctuations not only shape the internal dynamics of the halo but also provide a unique signature distinguishing FDM from CDM. The fluctuations act as a heating source, imparting random velocity kicks to stars and dark matter particles, leading to observable effects such as the thickening of stellar discs (Church, Mocz & Ostriker 2019; El-Zant et al. 2020), the broadening of tidal streams (Amorisco & Loeb 2018; Dalal et al. 2021), and the suppression of dynamical friction, which can prevent the inward migration of black holes and star clusters (Bar-Or, Fouvry & Tremaine 2019). Tidal fields from the host halo, such as the Milky Way, can suppress FDM-induced heating in satellites and reduce the tension between observations and predictions regarding the FDM particle mass (Schive et al. 2020; Yang et al. 2025).

The computational requirements for simulating full FDM dynamics have resulted in significant limitations in the cosmological volumes that can be studied (Woo & Chiueh 2009; Schive et al. 2014a; Veltmaat, Niemeyer & Schwabe 2018; Mocz et al. 2020). While methods that do not account for the full wave dynamics have allowed for simulations with larger volumes, closer to those achievable with traditional *N*-body and smoothed particle hydrodynamics (SPH) approaches for CDM (Schive et al. 2016; Veltmaat & Niemeyer 2016; Nori & Baldi 2018; Zhang et al. 2018; Nori et al. 2019), these simulations fail to capture inherent wave phenomena such as interference effects. These wave effects can significantly influence the evolution of the system, especially on small scales (Li, Hui & Bryan 2019), raising concerns about the validity of results obtained without solving the fundamental wave equations.

In a study of the effects of granules on particle orbits, Dutta Chowdhury et al. (2021) used high-resolution simulations of an FDM halo of mass $6.6 \times 10^9 \, M_\odot$ with $m_b = 8 \times 10^{-23}$ eV), which was extracted from a larger cosmological volume. Dutta Chowdhury et al. (2021) explored a statistical, diffusion model to understand the cumulative impact of granules on particle orbits, and considered only point-like particles (such as stars or black holes), such that their results may not apply to objects with sizes comparable to the scale of the granules, such as subhaloes, in which the finite size may significantly influence the magnitude of stochastic perturbations.

In this study, we introduce a semi-analytic model for the impact of FDM granules on subhaloes that can overcome many of the limitations of numerical simulations. Instead of a diffusion-based approach, we represent the stochastic perturbations as a Fourier series with random coefficients, allowing us to study their effects on individual orbits rather than just statistical populations. Using numerical simulations from Dutta Chowdhury et al. (2021), we calibrate our model and compare it to the previous methods. We also incorporate finite-size effects to examine how an extended subhalo interacts with the granular FDM environment. To validate our finite-size model, we developed FDM-SIMULATOR, which is designed to study how subhaloes move and interact within a background density field. Our results provide a more detailed understanding of how FDM quantum fluctuations impact subhalo dynamics.

This paper is organized as follows. In Section 2, we briefly review the structure of the FDM halo. We also introduce the Paley–Wiener representation of Brownian motion and describe how it is applied to generate random walks due to granule perturbations. Section 3 describes the results of our point-mass model. In Section 4, we describe FDM-SIMULATOR, which is designed to test and validate our model for perturbations of finite-size subhaloes, and introduce our results from the simulation and our semi-analytic model. We summarize and conclude in Section 5. Additionally, we compare our model with theoretical expectations from kinetic theory in Appendix A. Appendix B presents a detailed analysis of the statistical properties of granule-induced stochastic forcing, including the acceleration variance and the anisotropy of the resulting velocity dispersion.

## 2 METHODS

### 2.1 Fuzzy dark matter halo structure

FDM haloes consist of a central core embedded in an NFW-like density profile. The central core is a stationary, minimum-energy solution of the SP equation, called a 'soliton'. Quantum wave interference within an FDM halo generates persistent de Broglie-scale density fluctuations, known as granules. These granules are stochastically distributed and remain undamped over cosmological time-scales. As a result, the mass density profiles of dark matter haloes exhibit fine-scale fluctuations due to wave interference (Du et al. 2017; Hui et al. 2017). These granulations impart a markedly different structure to FDM haloes compared to those predicted by cold dark matter (CDM) and warm dark matter (WDM) models (Schive et al. 2014a; Schive et al. 2016; May & Springel 2023).

Outside the soliton, the density fluctuations within an FDM halo can be conceptualized as a sea of quasi-particles possessing an effective mass of

$$m_{\text{eff}} = \rho (f \lambda_{\text{db}})^3, \tag{1}$$

where $f = 0.282$ for a Gaussian velocity distribution in the halo particles (Bar-Or et al. 2019; El-Zant et al. 2020; Chiang, Schive & Chiueh 2021; Dutta Chowdhury et al. 2021), $\lambda_{\text{db}}$ is the de Broglie wavelength, which is given by

$$\lambda_{\text{db}} = \frac{\text{h}}{m_b \sigma_{\text{Jeans}}}, \tag{2}$$

where h is Planck's constant, $m_b$ is the mass of the FDM particle, and $\sigma_{\text{Jeans}}$ is the velocity dispersion in the halo found by solving the Jeans equation. In an FDM halo, self-gravity is counterbalanced by both random motion and quantum pressure (as discussed in Hui et al. 2017). When solving the Jeans equation, this results in an 'effective' velocity dispersion that is greater than the actual dispersion caused solely by random motion. $\sigma_{\text{Jeans}}$ is obtained by solving the spherical Jeans equation under the assumption of velocity isotropy with the time-averaged density profile of the halo. The velocity dispersion of quasi-particles is the same as the time-and-shell-averaged velocity dispersion $\sigma_h$, which is approximately equal to $\sigma_{\text{Jeans}}/\sqrt{2}$ (Dutta Chowdhury et al. 2021).

The effective mass of the quasi-particles can be translated into an effective size by equating $m_{\text{eff}}$ to $\rho V$, where $V = (\pi/6)d_{\text{eff}}^3$ is the volume of a quasi-particle of diameter $d_{\text{eff}}$, which is given by $d_{\text{eff}} \approx 0.35\lambda_{\text{db}}$ (Dutta Chowdhury et al. 2021). From this, we can estimate that the magnitude of the acceleration due to a nearby quasi-particle will be

$$a_{\text{eff}}^{\text{outskirt}} \approx \frac{Gm_{\text{eff}}}{d_{\text{eff}}^2}. \tag{3}$$

However, this equation is valid only outside the soliton. It breaks down inside the solitonic core, where the wavefunction is more coherent and the density profile is approximately flat. In this regime, we adopt a simplified prescription in which the outskirt contribution is fixed to a constant value equal to $a_{\rm eff}^{\rm outskirt}(r_{\rm sol})$ for $r < r_{\rm sol}$, where $r_{\rm sol}$ denotes the boundary of the soliton, beyond which the density profile transitions to an NFW-like behaviour.

To capture the full structure of the acceleration perturbations, we supplement the outskirt contribution with a separate component arising from the motion of the solitonic core. We define a characteristic core acceleration based on the enclosed solitonic core mass as

$$a_{\rm eff}^{\rm core} = \frac{GM(r_{\rm c})}{r_{\rm c}^2}, \qquad (4)$$

which is treated as spatially uniform within the solitonic core (i.e. for $r \lesssim r_{\rm c}$, where $r_{\rm c}$ is the core radius of the soliton) and set to zero for $r > r_{\rm c}$.

The frequency of the oscillations of the central soliton, which was first derived by Guzmán & Ureña-López (2004), is given by (equation 4 in Dutta Chowdhury et al. 2021, see Veltmaat & Niemeyer 2016 for more details):

$$f_{\rm granule}^{\rm soliton} = 10.94\,{\rm Gyr}^{-1}\left(\frac{\rho_0}{10^9\,{\rm M}_\odot\,{\rm kpc}^{-3}}\right)^{1/2}, \qquad (5)$$

where $\rho_0$ is the central density of the FDM halo. This frequency is derived from the quasi-normal mode of the ground state. In contrast, the excited states outside the soliton exhibit higher oscillation frequencies. An approximate estimate of the frequency of the granules in the outskirts can be obtained as follows. We expect the characteristic frequency to scale roughly as $f_{\rm granule} \sim \sigma_{\rm h}/d_{\rm eff}$. Substituting these expressions, we have

$$f_{\rm granule}^{\rm outskirt} = \frac{m_{\rm b}\sigma_{\rm Jeans}^2}{0.35\sqrt{2}\hbar}. \qquad (6)$$

To verify this result for the frequencies of granules, we run two simulations: (1) an isolated soliton with a central density of $10^8\,{\rm M}_\odot/{\rm kpc}^3$ and (2) a homogenous density field with a velocity dispersion of $17.8\,{\rm km\,s}^{-1}$ to mimic the outskirts of an FDM halo (full details of our simulation methodology are given in Section 4.1). We then compute the spectral density as

$$\mathcal{F}(f) = \frac{1}{V}\int\frac{1}{t_{\rm max}}\left|\int_0^{t_{\rm max}}\rho({\bf x},t)e^{i2\pi ft}{\rm d}t\right|^2{\rm d}{\bf x}^3. \qquad (7)$$

The spectral densities from these two simulations are shown in Fig. 1. For the case of a soliton (blue curve), the first peak of the spectrum corresponds to the frequency given by equation (5) (blue vertical line). For the homogenous case, the spectral density has a bell shape [1] We can define a characteristic spectral width, $f_0$, as $\sqrt{\mathcal{F}(f_0)/\mathcal{F}(0)} = 1/2$ (vertical orange line). $f_0$ is close to the frequency given by equation (6) (black vertical line). We note that while the wavefunction of the granules has a well-defined peak in its spectral density, the density (which is related to the inner product of the wavefunction with itself) does not, due to the non-linear coupling of modes. Nevertheless, the frequency given by equation (6) provides a useful estimate of the typical frequency of oscillations. We further note that had we computed the spectral density of the acceleration field due to granules (instead of their density field), we would have obtained a different estimate of $f_0$. We find that the precise choice of $f_{\rm granule}^{\rm outskirt}$ does not significantly affect our results.

---

[1] For this case, we subtract the mean density before computing the spectral density.

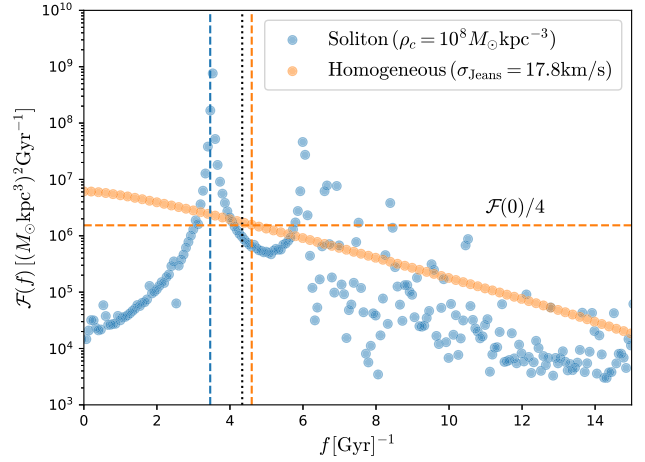

**Figure 1.** Spectral density of the solitonic core and homogeneous field. The blue vertical line shows the frequency of the core oscillation given by equation (5). The orange vertical line shows the characteristic spectral width of the homogenous density field, $f_0$, defined as $\sqrt{\mathcal{F}(f_0)/\mathcal{F}(0)} = 1/2$. $f_0$ is close to the value given by equation (6) (black vertical line).

### 2.2 Using the Paley–Wiener representation of a random walk

We consider that granules act as a source of stochastic gravitational accelerations, changing in both time and space, and that these perturbations will cause an orbiting body (e.g. a star, black hole, or subhalo) to experience a random walk in velocity. We use the Paley–Wiener representation of Brownian motion (equation 8 in Higham 2015), in which the coefficients of a Fourier series are random variables drawn from a normal distribution with zero mean. This allows us to generate a random walk in the velocity of the particle as a function of time:

$$v_x(\tau) = v_{x,0}\frac{\tau}{\sqrt{2\pi}} + \frac{2}{\sqrt{\pi}}\sum_{n=1}^m v_{x,n}\frac{\sin(n\tau/2)}{n} \qquad (8)$$

(for the $x$-component of velocity, with similar expressions for the $y$ and $z$-components), where $v_{x,i}$ are the random coefficients, $\tau$ is a dimensionless time that runs from 0 to $2\pi$, defined as

$$\tau = 2\pi\frac{t}{t_{\rm max}}, \qquad (9)$$

where $t_{\rm max}$ is the maximum time to which we want to run our calculation, and we choose $m$ such that the frequency of the fastest mode in the Paley–Wiener expansion equals the frequency of the granule structure oscillations[2]:

$$2\pi = \frac{m}{4}\tau_{\rm granule}, \qquad (10)$$

where $\tau_{\rm granule}$ is the oscillation period of the granules in the dimensionless time, given by $\tau_{\rm granule} = 2\pi/(f_{\rm granule}t_{\rm max})$. We then find $m$ using

$$m = 4f_{\rm granule}t_{\rm max}, \qquad (11)$$

where $f_{\rm granule}$ is given by equation (5). The factor of 4 was chosen empirically and found to be sufficient to ensure convergence in the resulting acceleration statistics over the full duration $t_{\rm max}$. Since $m$ must be an integer we first compute $m$ using the above for our desired $t_{\max}$, and then increase $t_{\max}$ to make $m$ equal to the next largest integer.

---

[2] In the usual Paley–Wiener representation of Brownian motion, $m \to \infty$. Here, we truncate the series as the FDM granules have a natural frequency above which we expect no fluctuations.

To incorporate the effects of the perturbations into particle orbit integrations, using acceleration is more direct than velocity. Therefore, we take the derivative of equation (8) with respect to time, finding an acceleration:

$$\frac{\mathrm{d}v_x(\tau)}{\mathrm{d}\tau}\frac{\mathrm{d}\tau}{\mathrm{d}t} = \frac{\sqrt{2\pi}}{t_{\max}}v_{x,0} + \sum_{n=1}^{m}\frac{2\sqrt{\pi}}{t_{\max}}v_{x,n}\cos(n\tau/2). \quad (12)$$

This suggests that the standard deviation of the coefficients in the series could be written as

$$\frac{2\sqrt{\pi}}{t_{\max}}v_{x,n} = a_{\mathrm{eff}} \approx \frac{Gm_{\mathrm{eff}}}{d_{\mathrm{eff}}^2}, \quad (13)$$

such that the coefficients in the terms in this acceleration series will also be drawn from a normal distribution of zero mean, but with a width

$$\sigma_a = a_{\mathrm{eff}}. \quad (14)$$

In practice, the stochastic acceleration field is composed of two physically distinct and statistically independent components: one from the solitonic core and one from the outer halo. These are modelled as separate Fourier series, each sampled from a normal distribution with width

$$\sigma_a^{\mathrm{core}} = \frac{1}{\sqrt{3}}a_{\mathrm{eff}}^{\mathrm{core}}, \quad \sigma_a^{\mathrm{outskirt}} = a_{\mathrm{eff}}^{\mathrm{outskirt}}, \quad (15)$$

where $a_{\mathrm{eff}}^{\mathrm{core}}$ and $a_{\mathrm{eff}}^{\mathrm{outskirt}}$ are defined in equations (4) and (3), respectively. These two components are generated independently and subsequently summed to yield the total stochastic acceleration field. The factor $1/\sqrt{3}$ in the core component arises because the solitonic-core-induced acceleration is modelled as isotropic in three dimensions, and $a_{\mathrm{eff}}^{\mathrm{core}}$ corresponds to the total RMS amplitude. In contrast, $a_{\mathrm{eff}}^{\mathrm{outskirt}}$ corresponds to the RMS value along a single direction and thus does not require the $1/\sqrt{3}$ normalization applied in the isotropic core case.

We can then write the series for the acceleration due to the effects of granules (for the $x$-component of acceleration, with similar expressions for the $y$ and $z$-components) as

$$a_x^{\mathrm{granule}}(\tau) = \frac{1}{\sqrt{f_{\mathrm{granule}}t_{\max}}}\left(\frac{1}{\sqrt{2}}a_{x,0} + \sum_{n=1}^{m}a_{x,n}\cos(n\tau/2)\right), \quad (16)$$

where the coefficients $a_{x,n}$ are sampled independently for the core and outskirt components, using normal distributions with widths given by equations (14) and (15). The scaling factor $f_{\mathrm{granule}}$ is likewise component dependent: for the core, we adopt the constant value in equation (5); for the outskirt, $f_{\mathrm{granule}}$ is fixed to the soliton-boundary value for $r < r_{\mathrm{sol}}$, and follows equation (6) otherwise. The pre-factor $1/\sqrt{f_{\mathrm{granule}}t_{\max}}$ ensures that the injected stochastic variance is independent of $t_{\max}$, consistent with the expected fluctuation time-scales derived from a truncated Paley–Wiener expansion of Brownian motion. This procedure yields two independent acceleration fields, $a_{\mathrm{core}}^{\mathrm{granule}}$ and $a_{\mathrm{outskirt}}^{\mathrm{granule}}$, which are subsequently combined to compute the total stochastic acceleration.

In Appendix B, we use FDM-SIMULATOR to show that the acceleration variance due to granules is both velocity-dependent and anisotropic. To model these effects, we introduce velocity-dependent scaling factors, $A_{\parallel}$ and $A_{\perp}$, corresponding to the components of the acceleration parallel and perpendicular to the particle's instantaneous velocity. These factors encapsulate how the particle's motion influences the coherence time and magnitude of granule interactions.

Specifically, the outskirt acceleration $a_{\mathrm{outskirt}}^{\mathrm{granule}}$ in equation (16) is rescaled by dividing the parallel and perpendicular components by $\sqrt{A_{\parallel}}$ and $\sqrt{A_{\perp}}$, respectively. The modified acceleration becomes

$$\boldsymbol{a}_{\mathrm{outskirt,scaled}}^{\mathrm{granule}} = \frac{\boldsymbol{a}_{\mathrm{outskirt},\parallel}^{\mathrm{granule}}}{\sqrt{A_{\parallel}}} + \frac{\boldsymbol{a}_{\mathrm{outskirt},\perp}^{\mathrm{granule}}}{\sqrt{A_{\perp}}}, \quad (17)$$

where $\boldsymbol{a}_{\mathrm{outskirt},\parallel}^{\mathrm{granule}} = (\boldsymbol{a}_{\mathrm{outskirt}}^{\mathrm{granule}} \cdot \hat{v})\hat{v}$ and $\boldsymbol{a}_{\mathrm{outskirt},\perp}^{\mathrm{granule}} = \boldsymbol{a}_{\mathrm{outskirt}}^{\mathrm{granule}} - \boldsymbol{a}_{\parallel}$, and $\hat{v}$ is the unit vector in the direction of the particle's velocity. Note that this velocity-dependent rescaling applies only to $\boldsymbol{a}_{\mathrm{outskirt}}^{\mathrm{granule}}$, and is not used for the core-induced component. The scaling factors $A_{\parallel}$ and $A_{\perp}$ are functions of velocity magnitude $v$ and background velocity dispersion $\sigma_J$, as given by equations (B28) and (B29).

## 3 POINT MASS OBJECT

As previously mentioned, the granules are variations in the density arising from the quantum nature of FDM, which will cause random perturbations to subhalo orbits. Dutta Chowdhury et al. (2021) model the effects of these perturbations through direct simulation and as a diffusion process. Throughout that paper, they simulated the impact of the perturbations in a single FDM halo of viral mass $M_{\mathrm{vir}} = 6.6 \times 10^9 \mathrm{M}_{\odot}$. The FDM halo is extracted from a cosmological simulation with a bosonic dark matter particle mass of $m_{\mathrm{b}} = 8 \times 10^{-23}\mathrm{eV}$ and a uniform spatial resolution of $\Delta x = 122$ pc, which is sufficient to resolve both the soliton core and the rest of the halo (Schive et al. 2014b). Here, we apply our model for the acceleration derived from the Paley-Wiener representation of Brownian motion (equation 16) to model the effects of granule perturbations. To compare with the results of Dutta Chowdhury et al. (2021), we use the same density profile as in that paper, also represented in the top panel of Fig. 2. The density profile of the FDM halo is given by

$$\rho(r) = \begin{cases} \rho_{\mathrm{sol}}(r) = \frac{\rho_c}{(1+0.091(r/r_c)^2)^8}, & \text{for } r \leq r_{\mathrm{sol}}, \\ \rho_{\mathrm{nfw}}(r) = \frac{\rho_s}{(r/r_s)(1+r/r_s)^2}, & \text{for } r > r_{\mathrm{sol}}, \end{cases} \quad (18)$$

where $\rho_c$ is the central density and $r_c$ is the core radius, defined by $\rho_{\mathrm{sol}}(r_c) = \rho_c/2$. The core radius is $r_c = 0.72$ kpc, obtained from the scaling relation:

$$\left(\frac{r_c}{\mathrm{kpc}}\right)^4 = \frac{1.954 \times 10^9 \, \mathrm{M}_{\odot} \, \mathrm{kpc}^{-3}}{\rho_0}\left(\frac{m_{\mathrm{b}}}{10^{-23} \, \mathrm{eV}}\right)^{-2}, \quad (19)$$

which arises from the scaling symmetry of the SP equation (Seidel & Suen 1990; Guzmán & Ureña-López 2006). We consider $r_{\mathrm{sol}} = 2.7r_c$, which is the approximate radius at which the density profile begins to diverge from the soliton structure, transitioning to an NFW-like profile, at the boundary of the soliton. Recent studies, such as Chan et al. (2022), find that most haloes have transition radii $r_{\mathrm{sol}} \leq 3r_c$ and similarly, Yavetz, Li & Hui (2022) report even smaller values (e.g. $r_{\mathrm{sol}} \approx 2r_c$). Theoretically, ensuring a smooth transition from the solitonic core to an NFW profile requires continuity in both the density and its first derivative, implying that $r_{\mathrm{sol}} \leq 3r_c$ is required. Since the simulation model, we currently use for comparison follows the approach of Schive et al. (2014a), our choice of $r_{\mathrm{sol}} = 2.7r_c$ is consistent. Inside the soliton, quantum pressure dominates, while in the outer parts of the halo, gravity dominates and the formation of small-scale density fluctuations occurs (Chan et al. 2022). For the NFW-like density profile, the scale density $\rho_s = 5.5 \times 10^5 \, \mathrm{M}_{\odot}\mathrm{kpc}^{-3}$ can be obtained from the continuity condition for the density, with the scale radius $r_s = 10$ kpc (see Schive et al. 2014b; Dutta Chowdhury et al. 2021, for further details).

Using the density profile shown above, the de Broglie wavelength and the acceleration due to a nearby quasi-particle in this FDM halo can be computed by the method derived in Section 2.1, and are shown in the middle and bottom panels of Fig. 2, respectively. The middle panel presents the de Broglie wavelength $\lambda_{\rm db}$ as a function of $r$, given by equation (2). The blue curve shows how $\lambda_{\rm db}$ varies with distance from the centre, reflecting the quantum nature within the FDM halo. The bottom panel shows the outskirt contribution to the granule-induced acceleration, $a_{\rm eff}^{\rm outskirt}$, as described by equation (3). Within the soliton ($r < r_{\rm sol}$), this quantity is fixed to the boundary value $a_{\rm eff}^{\rm outskirt}(r_{\rm sol})$, reflecting the coherent nature of the core. At larger radii, $a_{\rm eff}^{\rm outskirt}$ decreases, and the influence of granules diminishes. In addition to this outskirt term, the solitonic core also contributes a constant core component, given by equation (4). Based on our simulations, we estimate the amplitude of this contribution to be approximately $a_{\rm eff}^{\rm core}(r \lesssim r_c) \sim 530.4$ km s$^{-1}$ Gyr$^{-1}$.

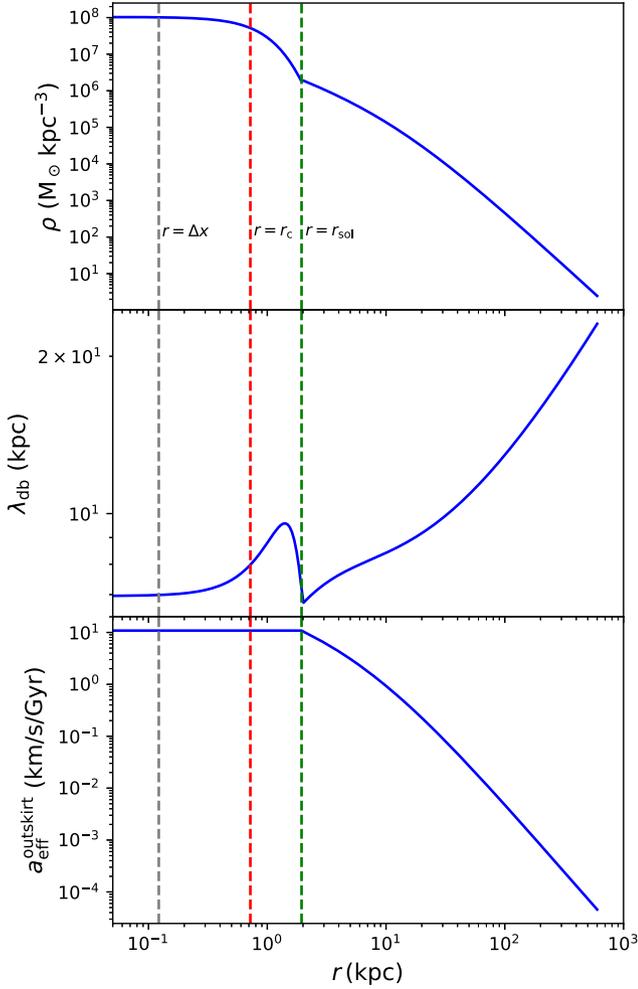

**Figure 2.** The structure of the FDM halo considered in this work. The top panel shows the density profile $\rho(r)$. The density profile exhibits a soliton-like structure at small radii, extending out to approximately $r_{\rm sol} = 2.7\,r_c$ (dashed vertical line at $r_{\rm sol}$), where $r_c = 0.72$ kpc is the core radius of the soliton (dashed vertical line at $r_c$). Beyond $r_{\rm sol}$, the profile transitions to an NFW-like structure at larger radii. The radius $r_{\rm sol}$ is considered to be the boundary of the soliton, marking the approximate point where the soliton ends and the outer halo begins. A third dashed vertical line marks the spatial resolution of the simulation in Dutta Chowdhury et al. 2021, $\Delta x = 122$ pc (i.e. 0.122 kpc), which is sufficient to resolve both the soliton core and the rest of the halo. Our calculations use analytic formulas and are not limited by this resolution. The middle panel shows the de Broglie wavelength $\lambda_{\rm db}$ as a function of radius $r$. The bottom panel presents the outskirt contribution to the acceleration, $a_{\rm eff}^{\rm outskirt}$, which is set to a constant value equal to $a_{\rm eff}^{\rm outskirt}(r_{\rm sol})$ within the soliton, and decreases beyond $r_{\rm sol}$.

To investigate the orbital evolution of an object within an FDM halo, we account for three sources of acceleration: first, the smooth acceleration,

$$a_{\rm smooth} = \frac{GM(r)}{r^2}, \qquad (20)$$

directed towards the centre of the halo, where $M(r)$ is the time-and-shell averaged enclosed mass at radius $r$; secondly, the acceleration due to the random perturbation as equation (16); and thirdly, the effect of dynamical friction, which we now describe.

Though FDM seems promising in explaining several small-scale observational inconsistencies with CDM theory, it can drastically change the dynamics of galaxies as compared to CDM. Some ways in which galactic dynamics in FDM vary most strongly from CDM are dynamical heating and friction due to the wave-like substructure of FDM. We consider the classical treatment of dynamical friction – see more details in Section 3.1 of Lancaster et al. (2020) and Dutta Chowdhury et al. (2021) – which was first derived by Chandrasekhar (1943).

Assuming the background to be an infinite medium composed of 'field' particles, whose distribution is homogeneous in position space and isotropic in velocity space, we model the cumulative effect of many two-body interactions between the field particles and the satellite mass, under the assumption that the satellite mass significantly exceeds that of the field particles. These approximations provide a framework for estimating the diffusion of the satellite or subject particle through phase space.

The dynamical friction coefficient, also known as the first-order diffusion coefficient parallel to the velocity of the particle $v$, is given by

$$D[\Delta v_\parallel] = -\frac{4\pi^2 G^2 M_{\rm particle} \rho \ln \Lambda_{\rm FDM}}{\sigma_{\rm Jeans}^2} \mathbb{G}(X), \qquad (21)$$

where $\rho$ and $\sigma_h$ are the time-and-shell-averaged density and velocity dispersion, which is approximately equal to $\sigma_{\rm Jeans}/\sqrt{2}$ (the same as that of the quasi-particles), $M_{\rm particle}$ is the mass of the subhalo, and $\mathbb{G}(X)$ is given by

$$\mathbb{G}(X) = \frac{1}{2X^2}\left[\mathrm{erf}(X) - \frac{2X}{\sqrt{\pi}}e^{-X^2}\right], \qquad (22)$$

where $X \equiv v/(\sqrt{2}\sigma_{\rm Jeans})$, and $\mathrm{erf}(X)$ is the error function. Here, the Coulomb logarithm is defined as

$$\ln \Lambda_{\rm FDM} = \ln \frac{4\pi r}{\lambda_{\rm db}}, \qquad (23)$$

where $\lambda_{\rm db}$ is de Broglie wavelength, given by equation (2), and $r$ is the distance from the soliton centre. The classical treatment of dynamical friction assumes that the medium through which the satellite travels is infinite and that the satellite is a point mass.

We can then use the above equations to relate the dynamical friction force, directed anti-parallel to the velocity of the orbiting body, to the diffusion coefficient as

$$F_{\rm DF} = \frac{8\pi^2 G^2 M_{\rm particle}^2 \rho \ln \Lambda_{\rm FDM}}{v^2} \mathbb{G}(X). \qquad (24)$$

Then the acceleration due to the dynamical friction is given by

$$\boldsymbol{a}^{\rm DF} = \frac{\boldsymbol{F}_{\rm DF}}{M_{\rm particle}}, \qquad (25)$$

which is directly opposite to the particle's velocity, resulting in a deceleration of its motion.

However, the 'standard' Chandrasekhar treatment is known to fail in reproducing the effect of stalling in cored density distributions (Read et al. 2006; Kaur & Sridhar 2018; Banik & van den Bosch 2021). To address this limitation, we improved our dynamical friction calculation by adopting the model introduced by Kaur & Sridhar (2018), which explicitly accounts for the core-stalling effect.

During the motion of the particle, we sum these three contributions to the acceleration: the smooth acceleration given by equation (20), the acceleration of the random perturbation, and the acceleration due to the dynamical friction given by equation (25). The granule-induced perturbations consist of two statistically independent components: one from the solitonic core and one from the outer halo. Both are generated using the Fourier series described in equation (16), with sampling widths determined by equation (15). The outskirt component is subsequently modified by a velocity-dependent rescaling following equation (17). Then the $x$-component of acceleration is given by

$$a_x = a_x^{\text{smooth}} + \frac{1}{\sqrt{3}}\alpha_{\text{core}} a_{\text{core},x}^{\text{granule}} + \alpha_{\text{outer}} a_{\text{outskirt,scaled},x}^{\text{granule}} + \beta a_x^{\text{DF}}, \quad (26)$$

where $\alpha_{\text{core}}$ and $\alpha_{\text{outer}}$ are dimensionless scaling parameters for the core and outskirt components, respectively. The value of $\alpha_{\text{core}}$ would be of order unity if the stochastic acceleration were dominated by the influence of a single granule, as expected in the coherent regime of the solitonic core. Outside the soliton, since multiple granules contribute, their accelerations partially cancel out. We therefore introduce a $\alpha_{\text{outer}}$ that accounts for the fact that an orbiting object interacts simultaneously with many granules, and we calibrate its value to match the results of numerical simulations. Similarly, $\beta$ is a dimensionless parameter to calibrate the strength of dynamical friction.

By integrating the above formula for acceleration over time, it is possible to investigate the orbital evolution of an object in an FDM halo. At $t = 0$, we initialize the simulation by placing a single particle at a radial distance $r_0 = 0.1$ kpc from the centre of the FDM halo, with an initial velocity determined by the circular velocity at that radius, given by $v_0 = \sqrt{GM_{r_0}/r_0}$, where $M_{r_0}$ is the enclosed mass at $r_0$, and directed perpendicular to the radial vector resulting in an initially circular orbit. The object is then evolved for a maximum time of 25 Gyr, using a time-step of $\Delta t = 0.001$ Gyr. At each time-step, the acceleration acting on the particle, given by equation (26), is computed as the sum of four components: (1) the smooth acceleration, (2) a stochastic acceleration from solitonic core fluctuations, (3) a velocity-dependent rescaling stochastic acceleration from outer-halo granules, and (4) the dynamical friction caused by interactions with the background field. Both stochastic components are initially modelled using a truncated Paley–Wiener representation of Brownian motion.

As described above, the acceleration due to the effects of granules is modelled using two statistically independent components: one from the solitonic core and one from the outer halo. Each is realized as a separate Fourier series, with coefficients $a_{x,n}$ drawn from normal distributions with zero mean and widths, $\sigma_a^{\text{core}}$ and $\sigma_a^{\text{outskirt}}$, given by equation (15). At each time-step, the values of $a_{\text{eff}}^{\text{core}}(r)$ and $a_{\text{eff}}^{\text{outskirt}}(r)$ are evaluated at the particle's current position, and used to compute the corresponding widths.

We sample the Fourier coefficients from normal distributions with zero mean and unit width, and scale them by the appropriate combination of $\alpha$ and $a_{\text{eff}}$: specifically, we use $\alpha_{\text{core}} a_{\text{eff}}^{\text{core}}/\sqrt{3}$ for the solitonic core component, and $\alpha_{\text{outer}} a_{\text{eff}}^{\text{outskirt}}$ for the outer halo component. This scaling must be applied at each time-step, as both $a_{\text{eff}}^{\text{core}}$ and $a_{\text{eff}}^{\text{outskirt}}$ can only be determined once the integration reaches that time-step.

At each time-step, we also determine the velocity-dependent factors $A_\parallel$ and $A_\perp$, using the particle's velocity $v$ and the background velocity dispersion $\sigma_J(r)$, according to equations (B28) and (B29). These are then used to rescale the parallel and perpendicular components of the acceleration, following equation (17), to incorporate the anisotropic response to granule perturbations observed in simulations. The object's position and velocity are updated iteratively using these accelerations, and its trajectory is tracked over time. This approach allows us to study granules' influence on the object's orbital evolution in an FDM halo.

We search for the parameters, $\alpha_{\text{core}}$, $\alpha_{\text{outer}}$, and $\beta$, which result in the best match with the simulation data from Dutta Chowdhury et al. (2021) by minimizing the figure-of-merit function:

$$\chi^2 = \sum_{i=1}^{N_{\text{steps}}} \left(\ln r_i - \ln r_i^{\text{sim}}\right)^2, \quad (27)$$

where $N_{\text{steps}}$ is the total number of time-steps in the simulation. At each time-step, we record the radius $r_i$ predicted by our model and compare it to $r_i^{\text{sim}}$, the corresponding value from the numerical simulation. Since we run multiple realizations, $r_i$ represents the mean radius over all realizations at time-step $i$.

The results for various particle masses are shown in Fig. 3. We find that $\alpha_{\text{core}} = 0.3$, $\alpha_{\text{outer}} = 5.0$, and $\beta = 0.16$ result in an excellent match of our semi-analytic model with the results of the numerical simulation. Notably, the adopted value $\alpha_{\text{outer}} = 5.0$ is consistent with the range inferred from our FDM-SIMULATOR analysis as we will show in Section 4.2.

In the left-hand panels of Fig. 3, from top to bottom, the solid curves with various colours show the evolution of the median distance from the soliton centre over time, calculated from 100 realizations, using our semi-analytic point-mass model, for masses {massless, $1 \times 10^5 M_\odot$, $1 \times 10^6 M_\odot$, $5 \times 10^6 M_\odot$, $1 \times 10^7 M_\odot$}. In comparison, the green lines show the results of the numerical simulation. In each case, the shaded region with the same colour highlights the 16$^{\text{th}}$–84$^{\text{th}}$ percentile variation in $r$.

In the absence of dynamical friction, massless particles experience continuous heating from FDM potential fluctuations, causing their orbital radius to increase over time on average. For massive particles, the radius increases until the heating rate from FDM fluctuations balances the cooling rate due to dynamical friction, reaching an equilibrium where the radius, $r$, and velocity dispersion, $\sigma$, stabilize.

For particles with masses $M = 10^5 M_\odot$ and $M = 10^6 M_\odot$, the heating rate dominates. In contrast, for $M = 5 \times 10^6 M_\odot$ and $M = 10^7 M_\odot$ dynamical friction becomes significant. For $M = 10^7 M_\odot$, the balance between heating and cooling results in a slight decrease in $r$ at late times in the simulations of Dutta Chowdhury et al. (2021). This decay is likely due to the gradual increase in the mass of the soliton over time through gravitational cooling. As the soliton gains mass, its random motion weakens, reducing the heating effect and causing the equilibrium radius to shrink slowly. Our model does not include this gravitational cooling effect, so we do not see the decrease in $r$ at late times.

The solid curves in the right-hand panels show the particle ensemble's 1D velocity dispersion, $\sigma$. Here, we compute the velocity dispersion in the $x$-direction (similar results are found for the velocity dispersion in the $y$ and $z$-directions):

$$\sigma_x = \langle(\Delta v)^2\rangle = \frac{1}{N}\sum_{i=1}^{N}(\Delta v_i)^2, \quad (28)$$

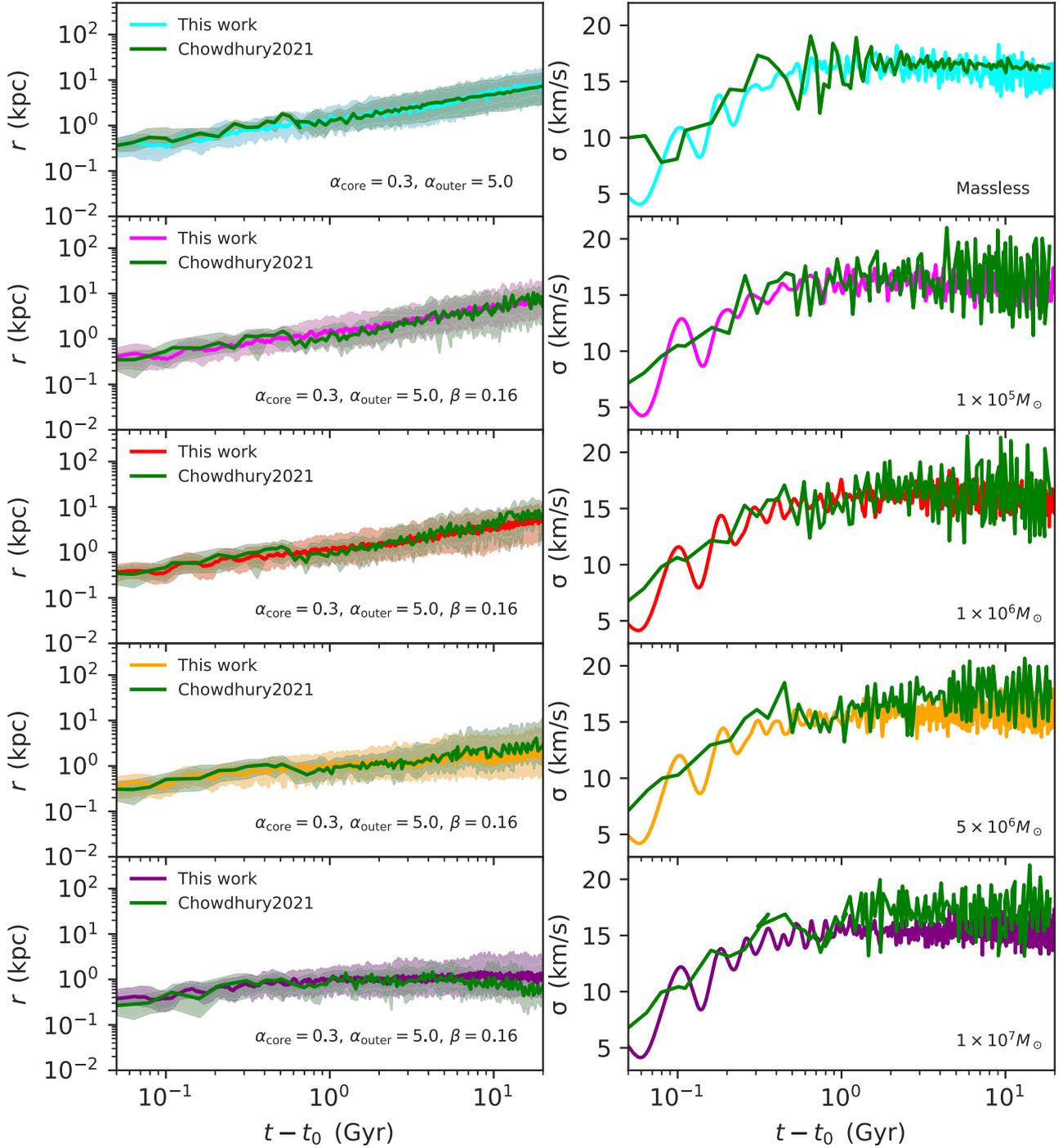

**Figure 3.** The orbital evolution of an object in an FDM halo. Each row corresponds to a different particle mass, from top to bottom: massless particles, $1 \times 10^5 M_\odot$, $1 \times 10^6 M_\odot$, $5 \times 10^6 M_\odot$, and $1 \times 10^7 M_\odot$ as indicate in the right-hand panels. In the left-hand panels, solid lines labelled 'This work' show the median radius of the particles over time, along with the 16th–84th percentile variation in radius, using our semi-analytic model for point-mass objects. For various particle masses, we calibrated a consistent scaling parameter, $\alpha_{\rm core} = 0.3$, $\alpha_{\rm outer} = 5.0$, to adjust the granule acceleration, and $\beta = 0.16$, to scale the dynamical friction effects, comparing our semi-analytic model with numerical simulation (lines labelled 'Chowdhury2021'). The right-hand panels display the velocity dispersion $\sigma$.

where $\Delta v_i = v_i - \bar{v}$, $v_i$ is the velocity over different realizations and $\bar{v}$ is the mean velocity. Then the velocity dispersion is $\sigma = \sqrt{(\sigma_x^2 + \sigma_y^2 + \sigma_z^2)/3}$. The evolution of $\sigma$ shows minimal mass dependence. Initially, $\sigma$ increases rapidly, as particles start with zero velocity relative to the soliton. After this phase, $\sigma$ evolves only slowly.

By parametrizing the granule acceleration and dynamical friction through constant $\alpha$ and $\beta$, we can account for a wide range of particle masses without overcomplicating the underlying framework. Our semi-analytic model integrates a particle orbit in around 1 minute, allowing for a much faster parameter space exploration than a full numerical simulation.

## 4 FINITE-SIZE OBJECTS

We have developed our semi-analytic model for point-mass objects. However, for extended objects, such as subhaloes, their spatial extent and structure may significantly influence the stochastic perturbations that they experience.

For an extended object, if its size exceeds the typical size and spacing of granules, the strength of perturbations from those granules that it encompasses will be significantly reduced – it will effectively smooth over some larger region containing many granules, resulting in the acceleration due to granules having a lower amplitude.

### 4.1 FDM Simulator

To validate that our finite-size model is physically accurate and reliable, we test it against direct FDM simulations. While the model captures the main statistical properties of granule-induced fluctuations, their interactions with extended objects can be more complex than analytic approaches assume. Simulations are therefore essential to account for the finite-size effects of FDM granules on extended subhaloes.

We simulate the motion of subhaloes within a background density field using FDM-SIMULATOR, a pseudo-spectral code for FDM simulations (Du et al. 2018). The initial background field is created as a superposition of random waves with a specified momentum distribution $f(k) \equiv |\psi(k)|^2$ (Bar-Or et al. 2019; Lancaster et al. 2020; Buehler & Desjacques 2023):

$$\psi(\mathbf{x}, 0) = \int d^3 \mathbf{k}\, \psi(\mathbf{k}) e^{i[\mathbf{k}\cdot\mathbf{x} + \phi(\mathbf{k})]}, \tag{29}$$

where $\phi(\mathbf{k})$ is a random phase uniformly drawn from $[0, 2\pi)$ for each $\mathbf{k}$ mode. The background density is $\rho(\mathbf{x}) = \rho_0 |\psi(\mathbf{x})|^2$ with $\rho_0$ being the mean density. It is found in cosmological simulations that in a virialized FDM halo $f(k)$ is roughly given by a Maxwell–Boltzmann distribution (Veltmaat et al. 2018)

$$4\pi k^2 f(k) dk = 4\pi v^2 f(v) dv$$
$$= 4\pi v^2 \left(\frac{1}{2\pi\sigma_{\text{Jeans}}^2}\right)^{3/2} \exp\left(-\frac{v^2}{2\sigma_{\text{Jeans}}^2}\right) dv, \tag{30}$$

where $v = \hbar k/m_b$, and $\sigma_{\text{Jeans}}$ is the velocity dispersion of the halo. Therefore, we also assume a Maxwell–Boltzmann distribution for the background field. The subhalo is modelled as a static NFW potential. Its acceleration is computed by integrating the acceleration from the background over this NFW profile:

$$\mathbf{a}(t) = \int \rho(\mathbf{x}, t) \frac{GM_{\text{NFW}}(|\mathbf{x} - \mathbf{x}_{\text{sub}}(t)|)}{|\mathbf{x} - \mathbf{x}_{\text{sub}}(t)|^2 + r_{\text{softening}}^2} \frac{\mathbf{x} - \mathbf{x}_{\text{sub}}(t)}{|\mathbf{x} - \mathbf{x}_{\text{sub}}(t)|} d^3 \mathbf{x}. \tag{31}$$

Here, $\rho$ is the background density, $M_{\text{NFW}}(r)$ is the enclosed mass of an NFW halo, $r_{\text{softening}}$ is the softening length, and $\mathbf{x}_{\text{sub}}$ is the position of the subhalo.

To study the effects of granules on the orbital evolution of the subhalo in an FDM halo, we focus exclusively on the acceleration caused by interactions with granules. For simplicity, we assume a uniform (apart from granule fluctuations) background such that there is no orbital acceleration, and ignore the effects of dynamical friction in the simulation (by disabling the gravitational back-reaction of the subhalo on the background matter). We also switch off the self-gravity of the background field to avoid possible nucleation in the background by kinetic relaxation (Levkov, Panin & Tkachev 2018; Chen et al. 2021).

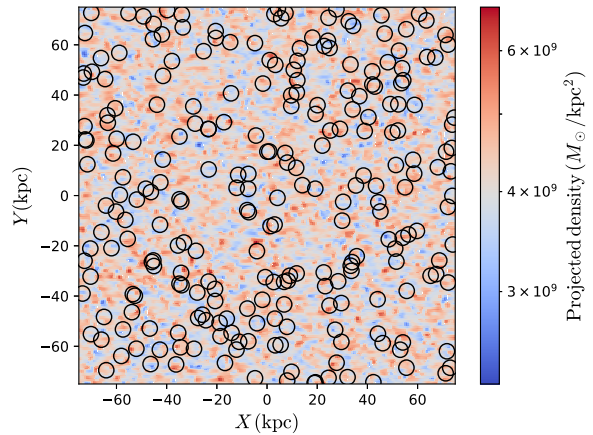

**Figure 4.** Projected background density field (colour map) with subhalo positions (black circles) in one realization using FDM-SIMULATOR. The granular structure of the FDM background is visible, highlighting the stochastic nature of the density fluctuations.

In the simulation, we can follow the motion of multiple subhaloes at once. These subhaloes are placed randomly in the background field and move around due to interactions with the granule structures in the field. Importantly, the subhaloes do not interact with each other – their motion is entirely driven by the background. By tracking many subhaloes simultaneously, we can measure statistical properties such as the root mean square velocity ($V_{\text{rms}}$) much faster than running individual simulations for each subhalo. However, because the simulation box is finite, the motion of subhaloes will be slightly correlated, and different realizations of the background field can lead to slight variations in the results. For example, the $V_{\text{rms}}$ can differ by about 7 per cent depending on the specific background field used. This variation is similar to that we find when changing the mass or size of the subhaloes, so it is essential to run multiple realizations to get reliable results. After testing, we found that running 20 realizations, each with 250 subhaloes, is enough to get consistent results for $V_{\text{rms}}$.

In the simulation, we assume a boson mass $m_b = 8 \times 10^{-23}$ eV. For the background, we use a host FDM halo with $M_{\text{host}} = 10^{10}$ M$_\odot$, which is a typical FDM halo mass scale, to estimate relevant densities. Specifically, we set the homogeneous density background as the value $\rho_0 = 1.628 \times 10^4$ M$_\odot$/kpc$^3$ at half of the virial radius, $r = 0.5 \times R_{\text{host}} = 28.35$ kpc of the host FDM halo, with the velocity dispersion set to $\sigma_{\text{Jeans}} = 17.775$ km s$^{-1}$, resulting in $d_{\text{eff}}$ set to 2.965 kpc. Using this same background field, we simulate subhaloes with various masses, from $3 \times 10^5$ M$_\odot$ to $9 \times 10^8$ M$_\odot$, associated with various half radii $r_{\text{half}}$, from 0.322 to 6.317 kpc, which is the radius enclosing half of the total mass of the subhalo. We simulate 250 subhaloes with the same mass in each realization and run 20 realizations for each subhalo mass.

In Figs 4 and 5, we show an example of the simulation results with $M_{\text{subhalo}} = 3 \times 10^6$ M$_\odot$. Fig. 4 shows the projected background density along with the positions of the 250 subhaloes (black circles) in one realization. In the top panel of Fig. 5, the blue curves show how the velocities of individual subhaloes change over time, with the black curve representing the root mean square velocity, $V_{\text{rms}}$, of the 250 subhaloes. The black dashed lines indicate the $\pm 1\sigma$ spread around $V_{\text{rms}}$. In the bottom panel, the blue curves show how the distance from the initial position, $r$, of individual subhaloes changes over time. The black solid line shows the median distance $r_{\rm med}$, while the black dashed lines indicate the 16th–84th percentile range.

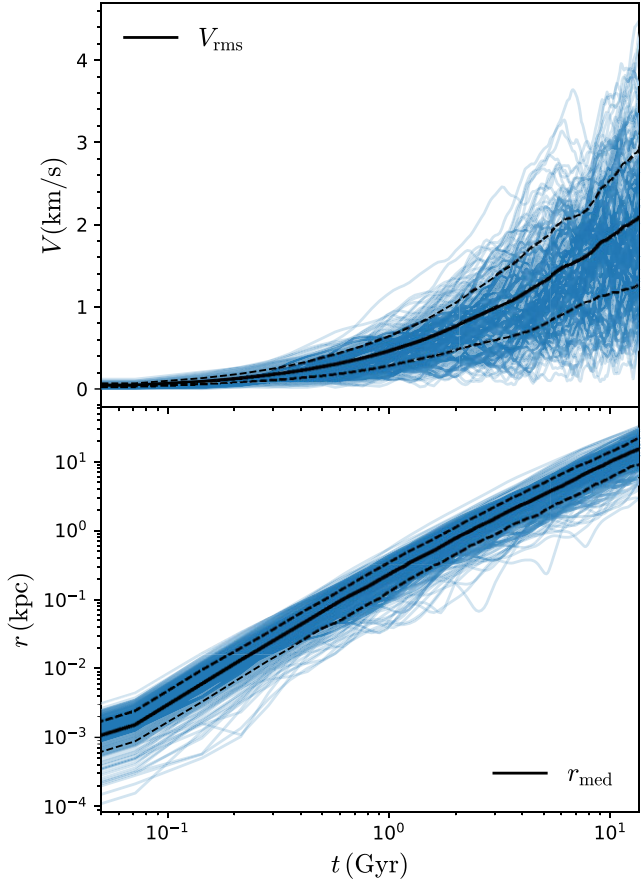

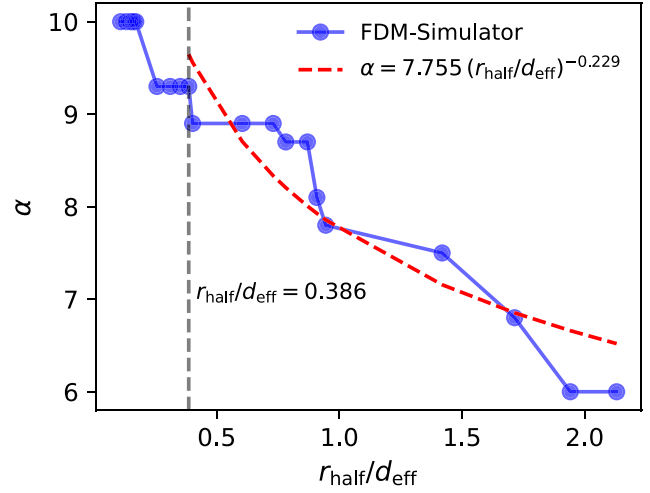

**Figure 5.** The time evolution of subhalo velocities and positions of the 250 subhaloes in a homogeneous FDM background in one realization, using FDM-SIMULATOR. In the top panel, the blue curves represent the velocity evolution of individual subhaloes over time, while the black curve shows the root mean square velocity ($V_{\rm rms}$), and the dashed curves mark the $\pm 1\sigma$ spread around it. In the bottom panel, the blue curves show the time evolution of the distance from the initial position, $r$, of individual subhaloes. The black solid line indicates the median distance ($r_{\rm med}$), while the black dashed lines mark the 16th–84th percentile range.

### 4.2 Finite-size model calibration with FDM simulations

In FDM-SIMULATOR, we consider only the effect on the motion of subhaloes due to the interaction with granules. Here, the $x$-component of acceleration in our finite-size model is given by

$$a_x = \alpha \, a_{{\rm scaled},x}^{\rm granule}, \tag{32}$$

where $\alpha$ is a dimensionless parameter of order unity that we calibrate to the results from FDM-SIMULATOR, and $a_{{\rm scaled},x}^{\rm granule}$, corresponding to the rescaled outskirt component, is given by equation (17). In this finite-size model, the quantity $a_{\rm eff}$, computed using equation (3), determines the amplitude of the stochastic acceleration.

We calibrate $\alpha$ to the results from FDM-SIMULATOR. As described above, in these simulations, subhaloes are modelled as NFW density profiles, and the acceleration they experience is computed by integrating over their mass distribution. For each subhalo mass, the mean value of $\alpha$ from 20 realizations is shown in Fig. 6. As expected, the value of $\alpha$ is related to the finite size of the subhalo, decreasing for larger subhaloes. In Fig. 6 we express this size using $r_{\rm half}$, the half-mass radius, calculated assuming an NFW profile, in units of $d_{\rm eff}$. The scale radius, $r_s$, in the NFW profile is determined using the concentration–mass relation (equation 9 in Dutton & Macciò 2014), which we apply to obtain the corresponding concentration for a given subhalo mass.

**Figure 6.** The scaling parameter, $\alpha$, of granule acceleration with subhalo half-radius $r_{\rm half}$, which is the radius enclosing half of the total mass of the subhalo. For small subhaloes, with $r_{\rm half}/d_{\rm eff} < 0.17$, $\alpha$ remains constant at 10.0, characteristic of point-mass behaviour. Between $0.26 \lesssim r_{\rm half}/d_{\rm eff} \lesssim 0.39$, $\alpha$ transitions to a lower but still point-like value of 9.3. For the subhaloes with $r_{\rm half}/d_{\rm eff} > 0.40$, $\alpha$ decreases with increasing subhalo mass. The decreasing trend is fit with a power-law formula: $\alpha = 7.755(r_{\rm half}/d_{\rm eff})^{-0.229}$. This trend reflects the suppression of granule effects in extended subhaloes due to their broader mass distribution.

In Fig. 6, as expected, in the small half-radius region where $r_{\rm half}/d_{\rm eff} < 0.17$ (corresponding to the low-mass range $M_{\rm subhalo} \lesssim 1 \times 10^6 \, {\rm M}_\odot$), we observe that $\alpha$ remains constant at $\alpha = 10.0$. Between $0.26 \lesssim r_{\rm half}/d_{\rm eff} \lesssim 0.39$ (corresponding to the low-mass range $3 \times 10^6 \, {\rm M}_\odot \lesssim M_{\rm subhalo} \lesssim 9 \times 10^6 \, {\rm M}_\odot$), $\alpha$ transitions to a lower value of $\alpha = 9.3$. In this regime, subhaloes perform as point mass particles and $\alpha$ does not depend significantly on subhalo size, consistent with previous results, as discussed in Section 3. This scale aligns with the theoretical lower bound on subhalo formation in FDM, where haloes or subhaloes smaller than $10^7 (m_{\rm b}/10^{-22}{\rm eV})^{-3/2} {\rm M}_\odot$ do not form (Hui et al. 2017). Given the bosonic dark matter particle mass used here, $m_{\rm b} = 8 \times 10^{-23}$ eV, subhaloes smaller than $\sim 1 \times 10^7 {\rm M}_\odot$ are not expected to form.

In the range $r_{\rm half}/d_{\rm eff} \in [0.40, 2.13]$ (corresponding to the mass range $M_{\rm subhalo} \in [1 \times 10^7 \, {\rm M}_\odot, 9 \times 10^8 \, {\rm M}_\odot]$), as $r_{\rm half}/d_{\rm eff}$ increases, there is a clear decreasing trend in $\alpha$. To quantify this decreasing trend, we fit the data in this range with a power-law formula:

$$\alpha = A \left( \frac{r_{\rm half}}{d_{\rm eff}} \right)^B, \tag{33}$$

where $A$ and $B$ are fitting parameters, $r_{\rm half}$ relates to the size of the subhalo, and $d_{\rm eff}$ is computed with the homogeneous background in the host FDM halo. The best-fitting parameters are found to be $A = 7.755$ and $B = -0.229$. This formula captures the decreasing trend of $\alpha$ with increasing subhalo half-mass radius, corresponding to the subhalo mass, providing a quantitative description of how $\alpha$ scales in this regime. The fit is consistent with the theoretical expectation that larger subhaloes experience reduced granule acceleration due to their extended mass distribution.

The values of $\alpha$ obtained from the finite-size model are systematically higher than the value $\alpha_{\rm outer} = 5$ found in our point-mass model, but remain within a comparable range. This modest difference likely reflects differences in modelling assumptions. In the finite-size simulation, both the background density and acceleration amplitude are held constant, corresponding to conditions at half the virial radius of the host halo. In the point-mass model, by contrast, the density profile declines with radius and $a_{\rm eff}$ decreases in the outer halo accordingly, as shown in the bottom panel of Fig. 2, reducing the contribution from distant granules.

The fixed background density and acceleration amplitude in the finite-size simulation enhance the influence of distant granules, leading to a slightly higher net acceleration that is absorbed into the fitted $\alpha$. Despite these differences, both models use the same velocity-dependent anisotropy through the $A$ factors calibrated from FDM-SIMULATOR, which helps maintain consistency in the overall range of $\alpha$ values.

Beyond this calibration, the framework may also be extended to account for finite-size effects more generally. While our results are based on subhaloes with NFW density profiles, the formalism presented in Appendix B allows one to compute the induced velocity dispersion for subhaloes with general internal mass distributions, provided the background acceleration field is sourced by FDM granules. Specifically, equation (B32) modifies the acceleration power spectrum to include the effect of the subhalo's density structure, and together with equation (B1), enables a numerical computation of the resulting velocity dispersion.

### 4.3 Granule effects on subhaloes

In the previous subsection, we derived a formula to predict $\alpha$ as a function of the half-radius of the subhalo, capturing how granule accelerations scale with subhalo mass. Based on this result, we now apply this formula to explore the relative importance of granule effects across a range of subhalo and host halo masses.

To quantify how granule acceleration affects the dynamics of subhaloes in FDM haloes, in Fig. 7 we present the ratio between granule acceleration (with $\alpha$ defined as in equation 33), and smooth acceleration at the soliton radius of the host halo, with a range of host halo mass from $10^{10} M_\odot$ to $9 \times 10^{10} M_\odot$, associated with the subhalo mass range from $10^7 M_\odot$ to $10^9 M_\odot$. This ratio decreases (meaning granule accelerations become less significant) as the subhalo or host mass increases.

The decline in the ratio with increasing subhalo mass occurs due to their larger half-mass radii, which reduces the accelerations from nearby granules. Similarly, the decrease with host halo mass arises because the smooth acceleration at the soliton radius scales with the soliton mass of the host halo and, therefore, with the host halo mass, while granule accelerations grow more weakly with halo mass. Thus, acceleration due to the smooth potential dominates in massive host haloes, diminishing the relative contribution of granules.

In general, the granule acceleration should have a few per cent effect on subhaloes, compared with the smooth acceleration. However, for lower mass subhaloes, the cumulative impact of granule accelerations may become significant over cosmological time-scales.

## 5 CONCLUSIONS

In FDM, the quantum nature of the axion-like dark matter field induces density variations within haloes, commonly referred to as 'granules'. These granules exert stochastic perturbations on the orbits of subhaloes, leading to their redistribution over time. Previous work has used numerical simulations or diffusion-based approaches to model these effects. In this study, we propose an alternative framework based on representing the perturbations as a Fourier series with random coefficients that apply to individual orbits, not just populations.

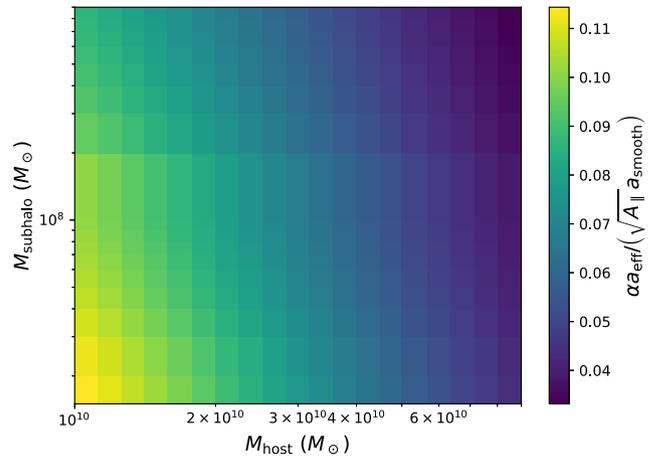

**Figure 7.** The colour map shows the ratio between the granule-induced acceleration (scaled by $\alpha$) and the smooth acceleration at the soliton radius of the host halo. This ratio, given by $\alpha a_{\rm eff}/\left(\sqrt{A_\parallel} a_{\rm smooth}\right)$, reflects the influence of the granule accelerations on subhalo dynamics, as a function of host halo mass $M_{\rm host}$ and subhalo mass $M_{\rm subhalo}$. Here, $a_{\rm eff}$ and $a_{\rm smooth}$ are the granule and smooth accelerations, respectively, both evaluated at $r_{\rm sol}$. $\alpha$ is the scaling factor accounting for finite-size effects, given by the equation we derived in the finite-size model. $A_\parallel$ is the anisotropic suppression factor of acceleration components aligned with the subhalo velocity, computed from the circular velocity and local velocity dispersion at $r_{\rm sol}$. The ratio tends to be highest for low-mass subhaloes in low-mass hosts, where granule-induced perturbations are most significant.

We summarize our results in the following points:

(1) **Point-mass particles:** We constructed a semi-analytic model for point-mass particles in FDM haloes, calibrating the scaling parameters $\alpha_{\rm core} = 0.3$ (core granule acceleration), $\alpha_{\rm outer} = 5.0$ (outskirts granule acceleration) and $\beta = 0.16$ (dynamical friction), based on simulations performed by Dutta Chowdhury et al. (2021). These parameters remain consistent across a wide range of particle masses, ensuring that our model can consistently reproduce results from numerical simulations and predict the granule acceleration for various particle masses.

(2) **Finite-size subhaloes:** Granules in FDM are inherently finite in size, and their spatial extent and structure can significantly influence the stochastic perturbations experienced by subhaloes. Unlike point-mass particles, finite-size subhaloes have an extended mass distribution, which modifies their response to granule accelerations. To account for this, we extended our model and derived a relation between the granule acceleration and the half radius, $r_{\rm half}$, of the subhalo.

We found that there is a distinct transition at $r_{\rm half}/d_{\rm eff} = 0.40$ (corresponding to $M_{\rm subhalo} = 1 \times 10^7 M_\odot$ with $m_b = 8 \times 10^{-23}$ eV) between the granule acceleration scaling parameter and subhalo mass. Below this mass scale, subhaloes behave like point-mass objects, as their compact size makes them highly sensitive to nearby granule fluctuations. However, for more massive subhaloes, the extended mass distribution reduces the effects of nearby granules, leading to weaker stochastic accelerations.

By deriving a relation between granule acceleration and $r_{\rm half}$, we provide a framework for understanding how finite-size effects influence subhalo dynamics in FDM haloes, which is useful to predict granule acceleration.

(3) **FDM-SIMULATOR**: To examine our finite-size model, we developed a simulation framework designed to validate the influence on subhalo dynamics of granule fluctuations in FDM haloes. The simulator represents subhaloes as static NFW potentials moving through a background density field made up of stochastic waves that mimic FDM granules. To focus on the effects of granule interactions, it isolates the resulting acceleration by removing contributions from dynamical friction and orbital motion. This ensures that the measured effects come specifically from the interaction between finite-size subhaloes and the granular background.

Specifically, the simulation uses a homogeneous background density field to isolate the effects of granule perturbations from other dynamical influences. In each run, multiple subhaloes (250 subhaloes per realization) are evolved simultaneously, with their motion driven entirely by stochastic interactions with the granular background. This setup enables efficient statistical analysis of key properties, such as the median value of distance $r$ from their initial positions. To account for variations from the finite simulation box size, we run 20 independent realizations for each subhalo mass. This helps reduce statistical noise and ensures consistent results. It also improves efficiency while revealing the impact of granule effects on the dynamics of subhaloes.

(4) **Granule effects on subhaloes:** We quantified the relative importance of granule accelerations by comparing them to smooth accelerations at the soliton radius of the host FDM halo. We found that the ratio between the granule acceleration and smooth acceleration decreases with both increasing subhalo mass and larger host halo mass. For more massive subhaloes, the extended mass distribution reduces the effects of granule fluctuations, making them less sensitive to stochastic perturbations. In more massive host haloes, the stronger smooth acceleration at the soliton radius dominates, diminishing the influence of granules on subhalo dynamics.

In conclusion, our study provides a comprehensive framework for understanding the effects of granule perturbations on subhalo dynamics in FDM haloes. By developing a semi-analytic model of point-mass particles and extending it to finite-size subhaloes, we have shown how granule accelerations scale with subhalo properties. We have validated our model with numerical simulations and quantified the importance of granule accelerations.

Our semi-analytic model circumvents key limitations of numerical simulations in resolving small-scale FDM halo dynamics, particularly the computational cost and resolution challenges in capturing granule effects. This efficient model makes it possible to simulate FDM populations, including the effects of granule fluctuations. These FDM populations can then be contrasted with another strong dark matter candidate, CDM, which behaves similarly to FDM on large scales but exhibits distinct differences in small-scale structure and dynamics. In future work, by combining predictions for populations of FDM subhaloes with observational data, we aim to derive constraints on the FDM particle mass, $m_{\rm b}$.

## ACKNOWLEDGEMENTS

We are grateful to Philip Mocz for helpful and constructive comments that improved this paper. This work benefited from discussions during group meetings. We thank Andrew Robertson, Ana Bonaca, and Sachithra Weerasooriya for useful discussions. XD acknowledges support from the National Science Foundation through Grants No. NSF-AST-1836016 and No. NSF-AST-2205100, and by the Gordon and Betty Moore Foundation through Grant No. 8548.

## DATA AVAILABILITY

The simulation data used in this paper will be made available upon request. The code used to run the simulations, FDM-SIMULATOR, is currently under active development. We plan to make it publicly available in the future as part of our commitment to open and reproducible research.

## APPENDIX A: DIFFUSION COEFFICIENT

Here, we compare our semi-analytic model with theoretical prediction based on kinetic theory. As mentioned in Section 2.1, the density fluctuations outside the soliton within an FDM halo can be conceptualized as a sea of quasi-particles. Here, the diffusion coefficients are derived by assuming an infinite, homogeneous sea of quasi-particles, in which the unperturbed trajectory of the object is a straight line. The evolution of the energy of a particle can be described by the energy diffusion coefficient, $D[\Delta E]$, see more details in Bar-Or et al. (2019), El-Zant et al. (2020), Lancaster et al. (2020), Chavanis (2021), and Dutta Chowdhury et al. (2021). The first order diffusion coefficient was given by equation (21). The second-order coefficient $D[\Delta E]$, which describes the effect of the granules, is given by

$$D[\Delta E]_{\text{2nd order}} = \frac{1}{2} D[(\Delta v_\parallel)^2] + \frac{1}{2} D[(\Delta v_\perp)^2], \tag{A1}$$

where $v$ is the velocity of the particle, $D[(\Delta v_\parallel)^2]$ and $D[(\Delta v_\perp)^2]$ are the second-order velocity diffusion coefficients parallel and perpendicular to $v$, given by

$$D[(\Delta v_\parallel)^2] = \mathcal{D} \frac{G(X_{\text{eff}})}{X_{\text{eff}}}, \tag{A2}$$

$$D[(\Delta v_\perp)^2] = \mathcal{D} \frac{\text{erf}(X_{\text{eff}}) - G(X_{\text{eff}})}{X_{\text{eff}}}, \tag{A3}$$

where

$$\mathcal{D} = \frac{4\sqrt{2}\pi G^2 \rho m_{\text{eff}}}{\sigma_h} \ln \Lambda_{\text{FDM}}, \tag{A4}$$

$m_{\text{eff}}$ is the effective mass of the quasi-particles, which is given by equation (1), and the Coulomb logarithm $\Lambda_{\text{FDM}}$ is given by equation (23). Here, $\rho$ and $\sigma_h$ are the time-and-shell-averaged density and velocity dispersion, the same as velocity dispersion of the quasi-particles, which is approximately equal to $\sigma_{\text{Jeans}}/\sqrt{2}$, $r$ is the distance from the soliton centre, and $\mathbb{G}(X)$ is given by equation (22), where $X_{\text{eff}} = v/(\sqrt{2}\sigma_h)$.

We consider a discrete random walk process to model the stochastic velocity perturbations caused by granule interactions. Each 'step' corresponds to a velocity kick imparted by a granule over its characteristic time-scale $t_{\text{granule}} = 1/f_{\text{granule}}$. The velocity change per step is $\Delta t \approx a_{\text{granule}} t_{\text{granule}}$, where $a_{\text{granule}}$ is the granule acceleration. After $m$ steps, the mean squared velocity change will therefore be $\overline{v^2} = m(a_{\text{granule}} t_{\text{granule}})^2$. In the continuum limit, this simplifies to

$$\overline{v^2} = a_{\text{granule}}^2 t_{\text{granule}} (m t_{\text{granule}}) = a_{\text{granule}}^2 t_{\text{granule}} t. \tag{A5}$$

For a diffusion process governed by a second-order diffusion coefficient $D[\Delta E]_{\text{2nd order}}$, the mean squared velocity change is defined as $\overline{v^2} = D[\Delta E]_{\text{2nd order}} t$. Therefore, we can make the correspondence, $a_{\text{granule}} = \sqrt{\frac{D[\Delta E]_{\text{2nd order}}}{t_{\text{granule}}}}$. This relation bridges our random walk model to the continuum diffusion framework.

To test consistency, we compute $a_{\text{granule}}$ from both our semi-analytic model and the diffusion coefficient derived from kinetic theory. These expressions have similar dependencies on the system's physical parameters, but differ in their normalizations, so the random walk coefficients in our Fourier-based framework can be parametrized similarly to our point-mass model. Based on the discussion in Section 2.2, we also use equation (16) as the series for the acceleration due to the effects of granules $a_x^{\text{granule}}(\tau)$ (for the $x$-component of acceleration, with similar expressions for the $y$ and $z$-components), where we can choose the $a_{x,n}$ coefficients by sampling random numbers from a normal distribution with mean zero and width as $\sigma_a = \sqrt{D[\Delta E]_{\text{2nd order}} t_{\text{granule}}}$.

However, the velocity diffusion coefficients, given by equations (A2) and (A3), are derived under the assumption that velocity diffusion is mainly caused by many weak interactions with quasi-particles. Because of this, these equations are only valid well outside the soliton. Near the soliton, this assumption no longer holds, since the soliton itself acts as a single strong quasi-particle dominating the perturbations. To resolve this issue, Dutta Chowdhury et al. (2021) introduce an 'effective model', in which the diffusion coefficients within a certain radius, $r_{\text{cut}}$, are replaced by their values at $r_{\text{cut}}$. They found through trial and error that setting $r_{\text{cut}} \approx 2.3 r_c$ produced good agreement with simulation results, especially for low-mass particles. For consistency and to compare with their results, we adopt the same 'effective model' and use their value of $r_{\text{cut}}$ in our analysis.

We then compute the orbital evolution of the particle in the FDM halo. In addition to the stochastic granule-induced perturbations, we also include the smooth acceleration $a_x^{\text{smooth}}$ and the dynamical friction term $a_x^{\text{DF}}$ in the equation of motion. During motion, the acceleration acting on the particle is given by

$$a_x = a_x^{\text{smooth}} + \alpha a_x^{\text{granule}} + \beta a_x^{\text{DF}}, \tag{A6}$$

where $\alpha$ and $\beta$ are the calibrated scaling factors. Here, we also consider two separate scaling parameters for the granule acceleration, $\alpha_{\text{core}}$ for $r \leq r_c$ and $\alpha_{\text{outer}}$ for $r > r_c$. Note that, here, we choose to make the transition between $\alpha_{\text{core}}$ and $\alpha_{\text{outer}}$ at $r_c$. This choice is consistent with the definition of $a_{\text{eff}}^{\text{core}}$ in equation (4), and reflects the transition between the inner region, where the soliton acts as a coherent perturber dominating the dynamics, and the outer region, where velocity diffusion is driven by cumulative weak interactions with quasi-particles.

In the left-hand panels of Fig. A1, the solid curves with various colours show the evolution of the median distance from the soliton centre over time, calculated from 100 realizations, using the diffusion method, for various particle masses, while the green lines show the results of the numerical simulation. The right-hand panels show the velocity dispersion.

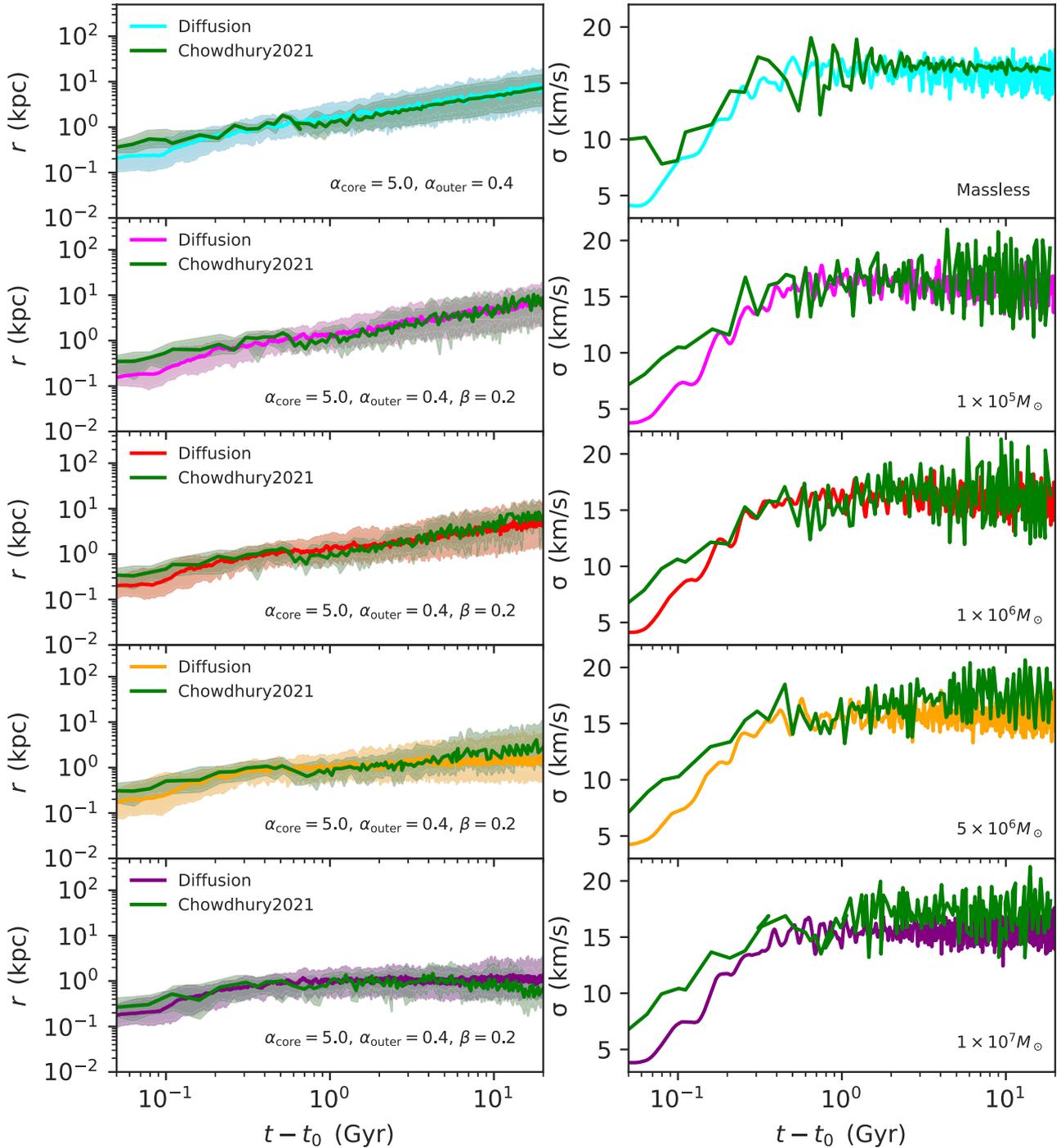

**Figure A1.** Same as Fig. 3, but using the diffusion model instead of our semi-analytic model to compute the granule-induced acceleration. In the left-hand panels, the solid lines with different colours show the median radius of the particles over time, using the diffusion method. For various particle masses, we calibrated a consistent scaling parameter, $\alpha_{\mathrm{core}} = 5.0$, $\alpha_{\mathrm{outer}} = 0.4$, to adjust the granule acceleration, and $\beta = 0.2$, to scale the dynamical friction effects. The right-hand panels display the velocity dispersion $\sigma$. The green lines show the results from the numerical simulation.

Using this diffusion approach, we find that the scaling parameters of granule acceleration are calibrated as $\alpha_{\mathrm{core}} = 5.0$ and $\alpha_{\mathrm{outer}} = 0.4$. Our value of $\alpha_{\mathrm{outer}}$ is much smaller than that we obtained in the point-mass model, reflecting the influence of the 'effective model'. We find the best scaling parameter of dynamical friction is $\beta = 0.2$.

## APPENDIX B: STATISTICAL PROPERTIES OF GRANULE FORCING

### B1 Velocity variance induced by granule fluctuations

Consider a point mass moving through a statistically homogeneous medium composed of FDM particles. Its velocity variance induced by the granule structure can be computed as (El-Zant, Freundlich & Combes 2016; Bar-Or et al. 2019; Marsh & Niemeyer 2019; El-Zant et al. 2020)

$$\sigma^2 \equiv \langle(\Delta v)^2\rangle = 2\int_0^t (t-t')\langle \mathbf{a}(\mathbf{x}=0, t=0)\mathbf{a}(\mathbf{x}=\mathbf{v}\,t', t')\rangle \mathrm{d}t'. \tag{B1}$$

Here, $\mathbf{a}(\mathbf{x}, t)$ is the acceleration field, and $\mathbf{v}$ is the velocity of the point mass. Therefore, $\sigma$ depends both on the spatial correlation and on the time correlation of the acceleration field, which can be defined as

$$C_a(\mathbf{x}, t) \equiv \langle \mathbf{a}(0, 0)\mathbf{a}(\mathbf{x}, t)\rangle. \tag{B2}$$

To compute the $C(\mathbf{x}, t)$, we assume the wavefunction of the homogeneous medium can be treated as a complex Gaussian random field:

$$\psi(\mathbf{x}, 0) = \int \mathrm{d}^3 \mathbf{k}\, \psi(\mathbf{k}) e^{i[\mathbf{k}\cdot\mathbf{x}+\phi(\mathbf{k})]}, \tag{B3}$$

where $\psi(\mathbf{k})$ is the Fourier transform of the wave function, and $\phi$ is a random phase. The evolution of the wavefunction is governed by the SP equations:

$$i\hbar\partial_t \psi(\mathbf{x}, t) = -\frac{\hbar^2}{2m_b}\nabla^2 \psi(\mathbf{x}, t) + m_b \Phi(\mathbf{x}, t)\psi(\mathbf{x}, t), \tag{B4}$$

$$\nabla^2 \Phi(\mathbf{x}, t) = 4\pi G \left(|\psi(\mathbf{x}, t)|^2 - \overline{\rho}\right), \tag{B5}$$

where $\Phi$ is the gravitational potential and $\overline{\rho} \equiv \overline{|\psi|^2}$ is the mean density. Here, we have assumed periodic boundary conditions. For convenience, we introduce dimensionless quantities:

$$\tilde{x} = x\frac{m_b v_0}{\hbar}, \quad \tilde{t} = t\frac{m_b v_0^2}{\hbar}, \quad \tilde{\psi} = \psi \frac{\hbar}{m_b v_0^2}\sqrt{4\pi G}, \quad \tilde{\Phi} = \frac{\Phi}{v_0^2}. \tag{B6}$$

Here, $v_0$ is a characteristic velocity. The SP equations can be rewritten as

$$i\partial_{\tilde{t}}\tilde{\psi}(\tilde{\mathbf{x}}, \tilde{t}) = -\tfrac{1}{2}\tilde{\nabla}^2 \tilde{\psi}(\tilde{\mathbf{x}}, \tilde{t}) + \tilde{\Phi}(\tilde{\mathbf{x}}, \tilde{t})\tilde{\psi}(\tilde{\mathbf{x}}, \tilde{t}), \tag{B7}$$

$$\tilde{\nabla}^2 \tilde{\Phi}(\tilde{\mathbf{x}}, \tilde{t}) = |\tilde{\psi}(\tilde{\mathbf{x}}, \tilde{t})|^2 - \tilde{\overline{\rho}}. \tag{B8}$$

For simplicity, we will omit the tilde symbols in the dimensionless quantities in the subsequent calculations. On time-scales much shorter than the gravitational condensation time-scale (Levkov et al. 2018; Chen et al. 2021), the self-gravity of the homogeneous median can be ignored. The evolution of wavefunction in Fourier space can be solved analytically:

$$\psi(\mathbf{k}, t) = \psi(\mathbf{k}, 0) e^{-\frac{1}{2} i k^2 t}. \tag{B9}$$

The time correlation can then be written as

$$C_\psi(\mathbf{k}, t) = \langle \psi^*(\mathbf{k}, 0)\psi(\mathbf{k}, t)\rangle = P(k)e^{-\frac{1}{2}ik^2 t}, \tag{B10}$$

where $P(k)$ is the power spectrum of the wavefunction. Assuming the homogeneous median has a Maxwell–Boltzmann velocity distribution, $P(k)$ is given by

$$P(k) = \frac{\overline{\rho}}{(2\pi\sigma_J)^{3/2}} \exp\left(-\frac{k^2}{2\sigma_J^2}\right), \tag{B11}$$

where $\sigma_J$ is the $1-D$ velocity dispersion and $k = |\mathbf{k}|$. For a Gaussian power spectrum, the spatial and time correlation function of $\psi$ can be computed analytically by doing an inverse Fourier transform of equation (B10):

$$C_\psi(r, t) = \overline{\rho}\, e^{\frac{i\sigma_J^2 r^2}{2(1+i\sigma_J^2 t)}} \left(1 + i\sigma_J^2 t\right)^{-3/2}, \tag{B12}$$

where $r = |\mathbf{x}|$.

Having the correlation function for $\psi$, the density correlation function can be computed as

$$\begin{aligned}\langle \rho(0,0)\rho(\mathbf{x}, t)\rangle &= \langle \psi^*(0,0)\psi(0,0)\psi^*(\mathbf{x}, t)\psi(\mathbf{x}, t)\rangle \\ &= \langle \psi^*(0,0)\psi(0,0)\rangle \langle \psi^*(\mathbf{x}, t)\psi(\mathbf{x}, t)\rangle \\ &\quad + \langle \psi^*(0,0)\psi^*(\mathbf{x}, t)\rangle \langle \psi(0,0)\psi(\mathbf{x}, t)\rangle \\ &\quad + \langle \psi^*(0,0)\psi(\mathbf{x}, t)\rangle \langle \psi(0,0)\psi^*(\mathbf{x}, t)\rangle \\ &= \overline{\rho}^2 + \langle \psi^*(0,0)\psi(\mathbf{x}, t)\rangle \langle \psi(0,0)\psi^*(\mathbf{x}, t)\rangle \\ &= \overline{\rho}^2 + |C_\psi(\mathbf{x}, t)|^2. \end{aligned} \tag{B13}$$

Here, we have used Wick's Theorem and the property $\langle \psi(0,0)\psi(\mathbf{x}, t)\rangle = 0$. Defining $C_\rho(r, t) = \langle \rho(0,0)\rho(\mathbf{x}, t)\rangle - \overline{\rho}^2$ and plugging in equation (B12), we have

$$C_\rho(r, t) = \overline{\rho}^2 e^{-\frac{\sigma_J^2 r^2}{1+\sigma_J^4 t^2}} \left(1 + \sigma_J^4 t^2\right)^{-3/2}. \tag{B14}$$

Its Fourier transform with respect to the spatial coordinates is

$$C_\rho(k, t) = \frac{\overline{\rho}^2}{8\pi^{3/2}\sigma_J^3} \exp\left(-\frac{k^2\left(1+\sigma_J^4 t^2\right)}{4\sigma_J^2}\right). \tag{B15}$$

From the Poisson equation (B8), it is easy to find that

$$C_\Phi(k, t) \equiv \langle \Phi(\mathbf{k}, 0)\Phi(\mathbf{k}, t)\rangle = \frac{C_\rho(k, t)}{k^4}. \tag{B16}$$

Here, $\Phi(k, t)$ is the Fourier transform of $\Phi(\mathbf{x}, t)$. For acceleration $\mathbf{a} = -\nabla\Phi$, we have

$$C_a(k, t) \equiv \langle \mathbf{a}(\mathbf{k}, 0)\mathbf{a}(\mathbf{k}, t)\rangle = k^2 C_\Phi(k, t) = \frac{C_\rho(k, t)}{k^2}. \tag{B17}$$

Performing an inverse Fourier transform, we obtain the correlation function for the acceleration field

$$C_{\mathbf{a}}(r, t) = \frac{\sqrt{\pi}\, \overline{\rho}^2}{4\sigma_J^3 r} \mathrm{erf}\left(\frac{\sigma_J r}{\sqrt{1+\sigma_J^4 t^2}}\right). \tag{B18}$$

Plugging equation (B18) into equation (B1), we can get the velocity dispersion of the point-mass particle induced by the granule structure in the homogeneous medium. If the velocity of the test particle $\mathbf{v} = 0$, the integral (B1) can be computed analytically:

$$\begin{aligned}\sigma^2 &= 2\int_0^t (t - t') C_a(0, t') \mathrm{d}t' \\ &= \frac{\overline{\rho}^2}{\sigma_J^6}\left[1 - \sqrt{1+\sigma_J^4 t^2} + \sigma_J^2 t\, \mathrm{arcsinh}\left(\sigma_J^2 t\right)\right], \end{aligned} \tag{B19}$$

where arcsinh is the inverse of the hyperbolic sine function.

### B2 Velocity dispersion anisotropy

In Appendix B1, we derive the formula for the velocity dispersion of a point-mass particle induced by the granule structure. In reality, the induced velocity dispersion is not isotropic, which can be seen as follows. Assuming the particle moves in the $x$ direction, the velocity dispersions along different axes can be computed as

$$\sigma_x^2 = 2\int_0^t (t-t') C_{a_\parallel}(x=v\,t', y=0, z=0, t') dt', \tag{B20}$$

$$\sigma_y^2 = \sigma_z^2 = \int_0^t (t-t') C_{a_\perp}(x=v\,t', y=0, z=0, t') dt', \tag{B21}$$

where $a_\parallel$ ($a_\perp$) is the acceleration parallel (perpendicular) to the moving direction of the particle. Similar to equation (B17), we have

$$C_{a_\parallel}(k, t) = \frac{k_x^2}{k^4} C_\rho(k, t), \quad (B22)$$

$$C_{a_\perp}(k, t) = \frac{k_y^2 + k_z^2}{k^4} C_\rho(k, t). \quad (B23)$$

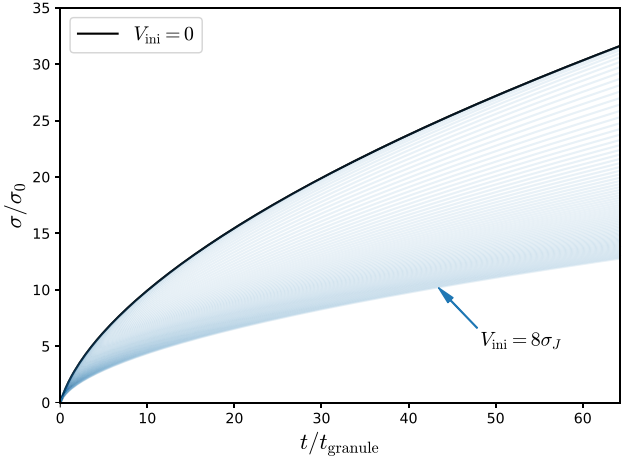

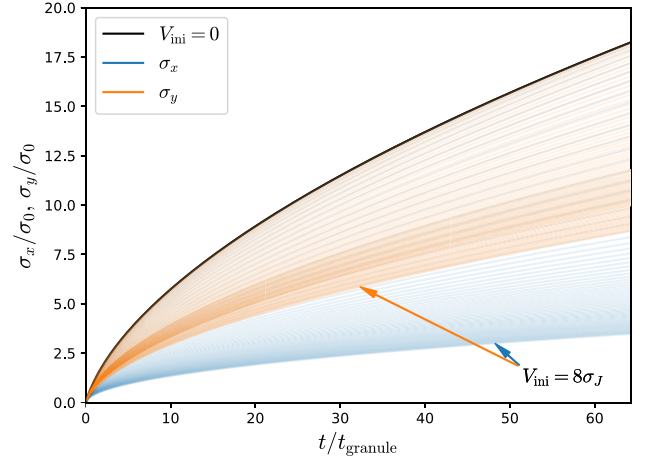

**Figure B1.** Time evolution of the velocity dispersion for particles with different initial velocities, shown in units of $\sigma_0 \equiv \bar{\rho}/\sigma_J^3$. The black curve shows the reference case with zero initial velocity ($V_{\rm ini} = 0$). As the initial velocity increases, the growth of the velocity dispersion becomes progressively slower.

**Figure B2.** Time evolution of the velocity dispersion for particles with nonzero initial velocities, shown in units of $\sigma_0 \equiv \bar{\rho}/\sigma_J^3$. The curves labelled '$\sigma_x$' and '$\sigma_y$' show the components parallel and perpendicular to the direction of motion, respectively. One curve corresponds to the reference case with zero initial velocity ($V_{\rm ini} = 0$), scaled by $1/\sqrt{3}$. The parallel component remains systematically smaller than the perpendicular one throughout the evolution, reflecting that the velocity dispersion becomes increasingly anisotropic with higher initial velocities.

We are only concerned with the spatial correlation along the moving direction ($x$-axis):

$$C_{a_\parallel}(x = vt, y = 0, z = 0, t)$$
$$= \iiint C_{a_\parallel}(k, t) e^{ik_x x} dk_x dk_y dk_z$$
$$= \frac{\bar{\rho}^2 \sqrt{1 + \sigma_J^4 t^2}}{4\sigma_J^4 x^2} \left[ -e^{-\frac{\sigma_J^2 x^2}{1+\sigma_J^4 t^2}} + \sqrt{\pi (1 + \sigma_J^4 t^2)} \frac{\mathrm{erf}\left(\frac{\sigma_J x}{\sqrt{1+\sigma_J^4 t^2}}\right)}{2\sigma_J x} \right], \quad (B24)$$

$$C_{a_\perp}(x = vt, y = 0, z = 0, t)$$
$$= \iiint C_{a_\perp}(k, t) e^{ik_x x} dk_x dk_y dk_z$$
$$= \frac{\bar{\rho}^2 \sqrt{1 + \sigma_J^4 t^2}}{4\sigma_J^4 x^2}$$
$$\left[ e^{-\frac{\sigma_J^2 x^2}{1+\sigma_J^4 t^2}} + \sqrt{\pi (1 + \sigma_J^4 t^2)} \left( -1 + 2\frac{\sigma_J^2 x^2}{1+\sigma_J^4 t^2} \right) \frac{\mathrm{erf}\left(\frac{\sigma_J x}{\sqrt{1+\sigma_J^4 t^2}}\right)}{2\sigma_J x} \right]. \quad (B25)$$

Plugging equations (B24) and (B25) into equations (B20) and (B21), we can see that $\sigma_x^2 \ne \sigma_y^2 = \sigma_z^2$.

Fig. B1 shows the total induced velocity dispersion $\sigma = \sqrt{\sigma_x^2 + \sigma_y^2 + \sigma_z^2}$ computed from equations (B1) and (B18) for different initial velocities. Note that for simplicity, we have fixed the particle's trajectory, i.e. the particle always moves along the $x$-axis with its initial velocity. In reality, the trajectory of the particle will deviate from the initial direction when it is kicked by the granules. As can be seen, particles with higher initial velocities experience a slower growth in velocity dispersion compared to those initially at rest.

In addition to the reduction in the dispersion growth rate, the velocity dispersion exhibits a clear anisotropy. Fig. B2 presents the time evolution of the velocity dispersion components parallel ($\sigma_x$) and perpendicular ($\sigma_y$) to the direction of motion computed from equations (B20), (B21), (B24), and (B25). For particles with non-zero initial velocities, the parallel component remains consistently smaller than the perpendicular one throughout the evolution.

We interpret this anisotropic behaviour in the velocity dispersion by introducing a velocity-dependent time rescaling factor $A$, which modifies the effective amplitude of the stochastic acceleration. Let $\sigma(t)$ denote the velocity dispersion for particles with zero initial velocity. For particles with nonzero initial velocity $v_{\rm norm}$, the time evolution of the velocity dispersion can be approximated as

$$\sigma(t; v_{\rm norm}) \sim \frac{\sigma(At)}{A}. \quad (B26)$$

Assuming $\sigma(t) \propto \sqrt{t}$, this leads to

$$\frac{\sigma(At)}{A} \propto \frac{\sigma(t)}{\sqrt{A}}, \quad (B27)$$

which implies that any rescaling of the time variable $t$ is equivalent to a modification of the amplitude of the acceleration. Based on this relation, we model the time rescaling factors for the total and perpendicular components of the velocity dispersion as

$$A_{\rm total} = \sqrt{1 + \left(\frac{v_{\rm norm}}{\sigma_J}\right)^2}, \quad A_\perp = \sqrt{1 + \frac{1}{2}\left(\frac{v_{\rm norm}}{\sigma_J}\right)^2}. \quad (B28)$$

The parallel component is then computed from

$$A_\parallel = \left(\frac{3}{A_{\rm total}} - \frac{2}{A_\perp}\right)^{-1}. \quad (B29)$$

These proposed expressions for the rescaling factors are validated by comparison with results extracted from semi-analytic calculations across a range of initial velocities. As shown in Fig. B3, the analytic forms for $A_{\rm total}$, $A_\perp$, and $A_\parallel$ provide excellent agreement with the extracted results.

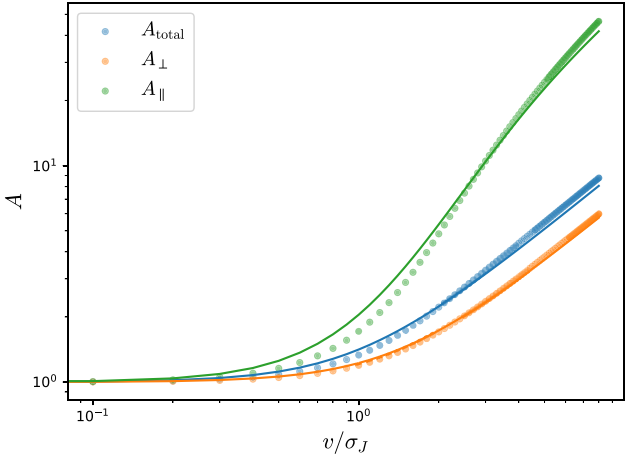
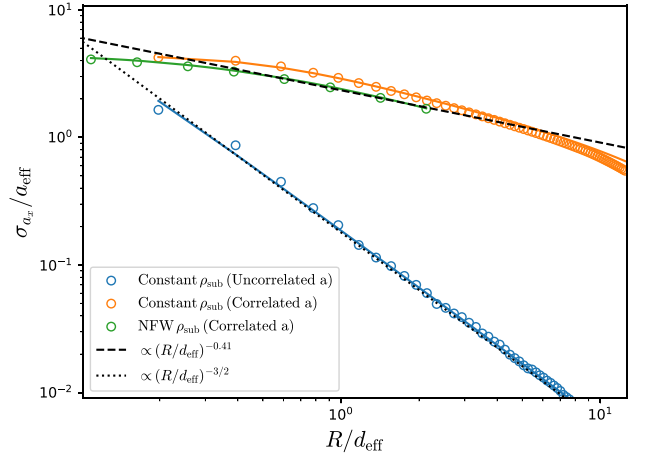

**Figure B3.** Velocity dependence of the rescaling factors $A_{\text{total}}$, $A_\perp$, and $A_\parallel$. The points show the values of each rescaling factor directly extracted from simulations with different initial velocities. The three sets of points correspond to $A_\parallel$ (upper set), $A_{\text{total}}$ (middle set) and $A_\perp$ (lower set). The solid curves are empirical fits to these data, following the relations: $A_{\text{total}} = \sqrt{1 + \left(\frac{v_{\text{norm}}}{\sigma_J}\right)^2}$, $A_\perp = \sqrt{1 + \frac{1}{2}\left(\frac{v_{\text{norm}}}{\sigma_J}\right)^2}$ and $A_\parallel = \left(\frac{3}{A_{\text{total}}} - \frac{2}{A_\perp}\right)^{-1}$.

**Figure B4.** Acceleration variance $\sigma_a$ as a function of subhalo size $R$, for different combinations of subhalo density profiles and acceleration field structures. Open circles show simulation results; solid curves are computed from the convolution of power spectra $P_{\rho_{\text{sub}}}(k)$ and $P_{\mathbf{a}}(k)$, where $P_{\rho_{\text{sub}}}(k)$ is the power spectra of the subhalo density profile and $P_{\mathbf{a}}(k)$ is that of the granule acceleration field. The blue points correspond to constant-density subhaloes in uncorrelated (white-noise) acceleration fields, showing Poisson scaling $\sigma_a \propto R^{-1.5}$. The orange points use the same subhalo profile but a granule-generated acceleration field, which is spatially correlated on the de Broglie wavelength scale. The green points represent NFW subhaloes in the same correlated field. For NFW subhaloes with sizes larger than $d_{\text{eff}}$, the acceleration variance decreases following a power law (black dashed curve).

## B3 Finite-size objects

We consider the granule-induced acceleration field $\mathbf{a}(\mathbf{x})$ as a statistically homogeneous Gaussian random field. The effective acceleration experienced by a subhalo, denoted $\mathbf{a}_{\text{sub}}$, is computed by convolving this field $\mathbf{a}(\mathbf{x})$ with the subhalo's density profile. Specifically, we evaluate the spatial average:

$$\mathbf{a}_{\text{sub}}(\mathbf{x}) = \frac{1}{M_{\text{sub}}} \int \rho_{\text{sub}}(|\mathbf{x}' - \mathbf{x}|)\,\mathbf{a}(\mathbf{x}')\,d^3\mathbf{x}', \quad (B30)$$

where $\rho_{\text{sub}}$ is the spherically symmetric density profile of the subhalo and $M_{\text{sub}}$ is its total mass. Under the assumption of Gaussian statistics, the variance of $\mathbf{a}_{\text{sub}}$ can be calculated in Fourier space as

$$\sigma^2_{\mathbf{a}_{\text{sub}}} = \int P_{\mathbf{a}_{\text{sub}}}(k)\,d^3\mathbf{k} = \frac{1}{M_{\text{sub}}^2} \int P_{\rho_{\text{sub}}}(k)\,P_{\mathbf{a}}(k)\,d^3\mathbf{k}, \quad (B31)$$

where $P_{\rho_{\text{sub}}}(k)$ and $P_{\mathbf{a}}(k)$ are the power spectra of the subhalo's density profile and that of the acceleration field, respectively. The variance is determined by the product of the subhalo's density power spectrum and the power spectrum of the acceleration field.

We test this approach using three representative cases, shown in Fig. B4. The open circles represent measurements extracted from the simulation output, while the solid curves correspond to theoretical predictions obtained from the convolution of the density and acceleration power spectra.

We consider constant-density subhaloes of varying radii, evaluated in an acceleration field with a white-noise power spectrum – i.e. spatially uncorrelated in structure, shown as the blue points in Fig. B4. In this case, the acceleration variance follows the expected Poisson scaling, $\sigma_a \propto N^{-1/2} \propto R^{-3/2}$, consistent with the fact that smaller volumes average over fewer independent realizations of the field. To isolate the effect of field correlations, we compute the acceleration variance using the same subhalo density profile but with an acceleration field generated from FDM granules, shown as the orange points. Unlike the white-noise case, this field is spatially correlated on the de Broglie wavelength scale, which leads to a larger acceleration variance compared to the uncorrelated case. The green points represent subhaloes with an NFW density profile, evaluated in the same acceleration field generated from FDM granules. The slightly lower variance observed for NFW subhaloes, particularly at small radii, is due to their centrally concentrated density profiles, which suppress low-$k$ modes in the density power spectrum. As a result, the NFW subhaloes are less sensitive to the large-scale acceleration fluctuations, leading to a reduced acceleration variance compared to constant-density subhaloes.

The theoretical predictions for the acceleration variance, $\sigma_{a_x}$, are in good agreement with the values measured from simulations across all three cases. For an NFW subhalo, $\sigma_{a_x}$ decreases with increasing $r_{\text{half}}$, following a power-law $\propto (r_{\text{half}}/d_{\text{eff}})^{-0.41}$. Note that, to facilitate comparison with simulation results, we impose lower and upper limits on the wavenumber when calculating the integral in equation (B31). Specifically, we set $k_{\text{min}} = \frac{2\pi}{\sqrt{3}L}$ and $k_{\text{max}} = \frac{\sqrt{3}\pi}{\Delta x}$ corresponding to maximum and minimum correlation lengths in a finite box, where $L$ is the size of simulation box and $\Delta x$ is the spatial resolution. Physically, if the subhalo is orbiting in a host halo of finite size, $k_{\text{min}}$ should be determined by the size of the host halo. This also explains why the coefficient $\alpha$ obtained by fitting to the homogenous background simulations (see Section 4.2) differs slightly from the value derived from the simulations of a realistic halo (see Section 3).

The induced velocity dispersion not only depends on the variance of the acceleration field, but also depends on its spatial and time correlation, as shown in Appendix B1. For finite-size objects, equation (B17) is modified as

$$C_{\mathbf{a}}^{\text{sub}(k,t)} = \frac{P_{\rho_{\text{sub}}}(k)}{M_{\text{sub}}^2}\,\frac{C_\rho(k,t)}{k^2}. \quad (B32)$$

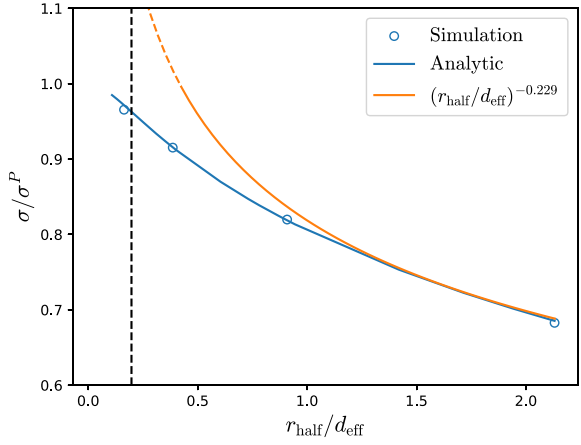

Unlike the point-mass case, the inverse Fourier transform of $C_a^{sub}(k, t)$ needs to be computed numerically. With the above equation, we can compute the induced velocity dispersion of an object with an arbitrary density profile.

Fig. B5 shows the velocity dispersion of a finite-size object relative to the point-mass case at $t = 50\, t_{granule}$. The analytic predictions (blue solid curve), computed from equations (B32) and (B1), agree well with the direct numerical simulations assuming NFW profiles for the subhaloes (circles). When the half-radius is larger than $d_{eff}$, the velocity dispersion decreases approximately as $(r_{half}/d_{eff})^{-0.229}$, consistent with the size dependence of the coefficient $\alpha$ (see Fig. 6).

**Figure B5.** Velocity dispersion of NFW subhaloes relative to the point-mass particle case at $t = 50\, t_{granule}$. The circles show the numerical simulation results, while the blue solid curve is computed from equations (B32) and (B1). The vertical dashed line marks the spatial resolution limit of the simulation. At $r_{half}/d_{eff} \gtrsim 1$, the ratio $\sigma/\sigma^P$ scales approximately as $(r_{half}/d_{eff})^{-0.229}$, consistent with the size dependence of the coefficient $\alpha$ obtained in Section 4.2.

This paper has been typeset from a T<sub>E</sub>X/LaT<sub>E</sub>X file prepared by the author.